\documentclass[journal,onecolumn,12pt]{IEEEtran}
  
\usepackage{graphicx}
\usepackage{amssymb}
\usepackage{amsmath}
\usepackage{amsthm}
\usepackage[linesnumbered,ruled]{algorithm2e}
\usepackage{mathrsfs}

\usepackage{epstopdf}%
\usepackage{cite} 
\newtheorem{prop}{Proposition}

\newtheorem{lemma}{Lemma} 
\usepackage{bbm}
\usepackage[bottom=0.5in,top=0.6in,left=0.6in,right=0.6in]{geometry}
\usepackage{color}

\def\AND{ }

\def\SMALL{ }

\begin{document}
 
\title{Secrecy Performance Analysis of UAV Transmissions Subject to Eavesdropping and Jamming}

\author{Jinchuan~Tang,~\IEEEmembership{Student~Member,~IEEE,}
        Gaojie~Chen,~\IEEEmembership{Member,~IEEE,}
        and~Justin~P. Coon,~\IEEEmembership{Senior~Member,~IEEE}
        \thanks{This work was supported by the EPSRC through ``Spatially Embedded Networks'' under Grant EP/N002350/1.}
\thanks{J. Tang and J. P. Coon are with the Department of Engineering Science, University of Oxford, Oxford, UK e-mail: \{jinchuan.tang, justin.coon\}@eng.ox.ac.uk.}
\thanks{G. Chen is with the Department of Engineering, University of Leicester, Leicester, UK e-mail: gaojie.chen@leicester.ac.uk.}
\thanks{This work has been submitted to the IEEE for possible publication. Copyright may be transferred without notice, after which this version
may no longer be accessible.}
}

\maketitle

\begin{abstract}
Unmanned aerial vehicles (UAVs) have been undergoing fast development for providing broader signal coverage and more extensive surveillance capabilities in military and civilian applications. Due to the broadcast nature of the wireless signal and the openness of the space, UAV eavesdroppers (UEDs) pose a potential threat to ground communications. In this paper, we consider the communications of a legitimate ground link in the presence of friendly jamming and UEDs within a finite area of space. 
The spatial distribution of the UEDs obeying a uniform binomial point process (BPP) is used to characterize the randomness of the UEDs. 
The ground link is assumed to experience log-distance path loss and Rayleigh fading, while free space path loss with/without the averaged excess path loss due to the environment is used for the air-to-ground/air-to-air links. 
A piecewise function is proposed to approximate the line-of-sight (LoS) probability for the air-to-ground links, which provides a better approximation than using the existing sigmoid-based fitting. The analytical expression for the secure connection probability (SCP) of the legitimate ground link in the presence of non-colluding UEDs is derived. 
The analysis reveals some useful trends in the SCP as a function of the transmit signal to jamming power ratio, the location of the UAV jammer, and the height of UAVs.
\end{abstract}


\IEEEpeerreviewmaketitle

\section{Introduction}

\IEEEPARstart{U}NMANNED aerial vehicles (UAVs), also known as drones, are a promising technology that offers reliable and cost-effective wireless communication solutions in a wide range of real-world scenarios \cite{mozaffari2018tutorial}. Recently, the low-altitude UAVs with elevated height from hundreds of meters to several kilometers have drawn much research attention in surveillance, public safety and secure communications \cite{8337903, naqvi2018drone, 8335329}. Compared to the
existing terrestrial communication systems, UAV aided wireless networks have the potential to overcome the propagation constraints due to terrain characteristics, enhance the signal coverage and reduce operating cost \cite{8335329}. Due to the broadcast nature of the wireless signal and the openness of the space, ground communications are susceptible to eavesdropping. UAV eavesdroppers (UEDs) could pose a greater threat to the security of ground communications than ground eavesdroppers, since UEDs are less constrained by terrain characteristics and a higher chance of line-of-sight (LoS) link with stronger signal strength can be formed from the ground transmitter to the UEDs rather than to ground eavesdroppers. Furthermore, many aspects of UAVs such as low production cost, high mobility, and ease of operation could incentivize the attackers to use UAVs as the major eavesdropping tools.  Consequently, it has become increasingly urgent and necessary to study the security of ground-based wireless communication in the presence of UEDs.

\subsection{Motivation and related work }
Among the many methods for securing wireless communication, physical layer security (PLS) has emerged as a promising technique for achieving a secure transmission by exploiting the channel characteristics through signal processing techniques and channel coding without the need of a shared secret key \cite{harrison2013coding, 7888950}. By using an information theoretic formalism, PLS has been shown to support perfect secrecy under realistic condition \cite{poor2017wireless}. Due to the fading effects of wireless channels and the unpredictable locations of eavesdroppers, probabilistic approaches have been used to characterize the likelihood of a link achieving a secrecy rate, namely, secure connection probability (SCP) \cite{yao2016secure}. Stochastic geometry has been exploited for the analysis of eavesdropping wireless networks by endowing the locations of the eavesdroppers with a probability distribution, such as the Poisson point process (PPP) or the binomial point process (BPP) \cite{pinto2012secure, chen2017secrecy, tang2018Meta}.  For example, the spectrum sharing of super dense drone small cell networks modeled by a 3D PPP is studied in \cite{zhang2017spectrum}. 
The authors of \cite{chetlur2017downlink} consider the case that the number of UAVs is small and deployed to cover a given finite region with a more suitable homogeneous BPP for UAV networks under Nakagami-m fading. 
Since it is common in deployment scenarios (especially in suburban and rural areas) to have a significantly stronger LoS component rather than reflected multipath components, the coverage probability in the absence of fading has been derived in \cite{chetlur2017downlink}. Furthermore, to capture the performance of an air-to-ground (ATG) link between a ground device and a UAV, the channel propagation model incorporating blockages from buildings is required. Based on the statistical model for building blockages \cite{itu2012Propagation}, the LoS probability in the product of a sequence of terms and its approximation via sigmoid function is formulated, and the optimal altitude for deploying the UAV with maximum coverage is studied in \cite{al2014optimal}.

There have been a lot of studies on either optimizing network resources or developing techniques for realistic system-level analysis of UAV networks on coverage, but it has been pointed out by \cite{8335329} that very few studies have investigated the secrecy performance of UAV networks, and only ground-based eavesdroppers were considered in those scenarios. In \cite{wang2017improving}, the optimization of the secrecy rate of a UAV-enabled mobile relay system was formulated, and a performance gain over static relaying was achieved. By jointly designing the trajectory and transmit power of the UAV, an optimization algorithm designed to achieve the average worst-case secrecy rate improvement in UAV-to-ground communications is proposed in \cite{8392472}. Additionally, \cite{8335329} investigated the secrecy performance of UAV networks working in the millimeter wave band, where the UAVs can be used either for information transmission or jamming, and it was revealed that the average achievable secrecy rate does not change monotonically with an increasing proportion of UAV jammers. Although it is known that increasing the jamming power will pose a stronger interference to both the UEDs and the legitimate ground receiver, how the secrecy performance behaves with respect to this increase is still unknown. Furthermore, the coverage and secrecy analysis in the previous literature has suffered from the tractability problem when the LoS probability is considered, since the analysis always results in multi-integral expressions.

\subsection{Contributions}
This paper focuses on the secrecy performance of ground-based communications in the presence of a UAV jammer and UEDs. The main contributions of the paper are summarized as follows:

\begin{enumerate}
\item {\bf LoS model:}  This is the first work to propose an approximation to simplify the modeling of the probability of LoS channels, which allows us to get tractable formulations in the analysis of wireless networks concerning ATG channels. The trend of the LoS probability with respect to the height of UAVs has been captured in a simplistic manner.

\item {\bf UAV jammer:} We introduce a UAV jammer for improving the security of the ground communication, and we give the cumulative distribution function (CDF) of the signal-to-interference-plus-noise ratio (SIR) from a ground transmitter to a UED subject to interference by the UAV jammer. 

\item {\bf SCP:} We formulate the SCP of the ground link in the presence of randomly deployed UEDs for the non-colluding scenario. The trends of the SCP in some environments with respect to different transmit signal to jamming power ratios, and the locations of the UAV jammer have been analysed. 

\end{enumerate}

The rest of the paper is structured as follows. Section II begins with a description of the system model then addresses the LoS probability and its approximations. Section III focuses on the derivation of the SCP. Section IV focuses on the behaviour of the SCP for different parameters. Section V gives the simulation results and discussion, and Section VI concludes the paper.

\section{System Model}
\subsection{Network layout}
\begin{figure} [t]
\centering
\includegraphics[width = 0.35 \textwidth]{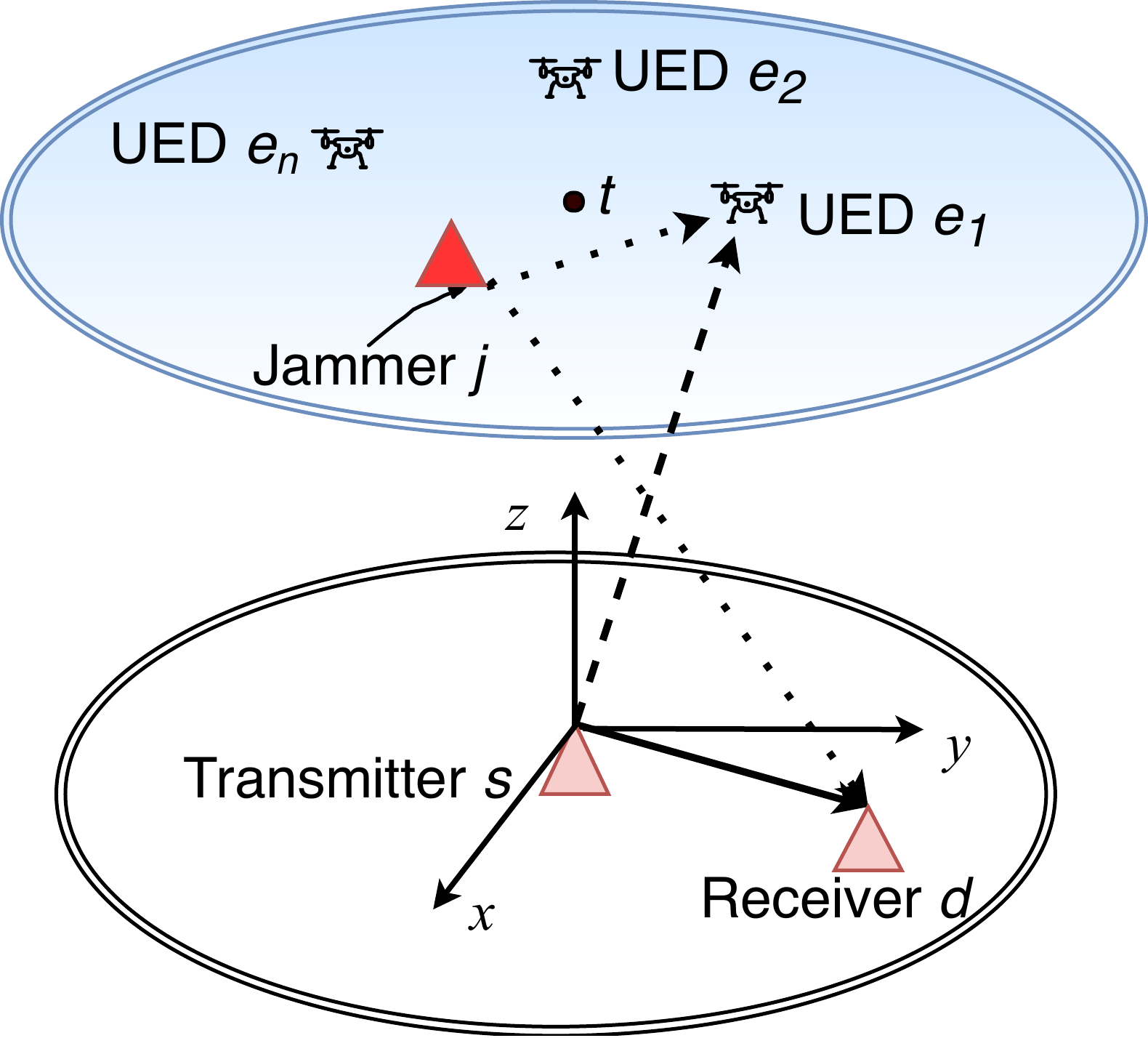}
\caption{The system model for communication in the presence of UEDs. The bold solid arrow denotes the legitimate link. For UED $e_1$, the dashed arrow denotes the wire-tap link, and the dotted arrows denote the jamming links.}
\label{UAV-fig}
\end{figure}

As shown in Fig. \ref{UAV-fig}, transmitter ${s}$ at $(0,0,0)\in \mathbb{R}^3$ connects to receiver ${d}$ at $(x_d,y_d,0)$ via the legitimate link, where $x_d$ and $y_d$ denotes the locations of $d$ on the x-axis and y-axis, respectively. There are $n$ UEDs uniformly distributed in a disk, forming a uniform binomial point process (BPP) \cite{chetlur2017downlink}. The disk is denoted as $O(t,R_{1})$, where $O$ denotes the two-dimensional disk of radius $R_{1}$ centred at $t$. The coordinates of the origin of the disk are $(0,0, H)$ so that the center of the circular plane $t$ is $H$ meters right above the transmitter $s$ on the ground, and the disk is also parallel to the ground \footnote{We assume that the ground between $s$ and $d$ is flat, and the effect due to the spherical surface of the earth is negligible. }. The UEDs work independently to decode the received signal on their links. Meanwhile, a UAV jammer $j$ is also located within the disk $O(t,R_{1})$ and continually sends a jamming signal.  
The ground channel is assumed to experience Rayleigh fading and path loss, while the channel model for the ATG communications is based on the probabilistic LoS and non-line-of-sight (NLoS) links given by \cite{al2014optimal, mozaffari2016unmanned}. The air-to-air communication channel is assumed to follow Friis free space transmission. The transmit powers of $s$ and $j$ are given by $P_s$ and $P_j$, respectively. 



\subsection{Channel model}
\subsubsection{Air-to-air channel model}
Since no major obstacles are obstructing the communications in the sky, the air-to-air channel follows Friis free space transmission. And the received jamming power from $j$ to $e$ is given by 
\begin{equation}
P_{j,e} = P_{j} (\lambda/4\pi)^\alpha l_{j,e}^{-2},
\end{equation}
where $\alpha =2$ is the free space path loss (FSPL) exponent, and $l_{{x},{y}}$ is the distance between nodes $x$ and $y$.

\subsubsection{ATG channel models}
The potential existence of buildings and other obstructions lying in the propagation path results in the presence of mixed LoS/NLoS channel conditions between an air terminal and a ground terminal. According to \cite{mozaffari2016unmanned}, the ATG channel model will be mainly based on probabilistic LoS and NLoS links instead of following a classical fading channel. Therefore, the corresponding received signal powers at $e$ from transmitter $s$ for LoS and NLoS links are written as

\begin{equation}
P_{s,e}=
\begin{cases}
 \eta_{\text{L}} P_s (\lambda/4\pi)^2 l_{s,e}^{-\alpha}, & \text{LoS link}\\ 
 \eta_{\text{N}}  P_s (\lambda/4\pi)^2 l_{s,e}^{-\alpha}, & \text{NLoS link},
\end{cases}
\end{equation}
and $\eta_{\text{L}}$ and $\eta_{\text{N}}$ refer to the mean values of excess path loss added to the FSPL, where $(\eta_{\text{L}},\eta_{\text{N}})$ can be measured at $f_c = 2$ GHz in dB to be $(-0.1,-21),(-1.0,-20),(-1.6,-23)$, and $(-2.3,-34)$ for suburban, urban, dense urban, and highrise urban areas, respectively \cite{al2014optimal}. 

\paragraph{LoS probability and sigmoid fitting}
The probabilities for a link $(s,e)$ to be either LoS or NLoS can be denoted as $\mathbb{P}_{\text{L}}^{s,e}$ and $\mathbb{P}_{\text{N}}^{s,e} =1 - \mathbb{P}_{\text{L}}^{s,e}$, respectively. According to \cite{itu2012Propagation, al2014optimal},
\begin{equation}\label{eq_LoS_NLoS}
\begin{split}
\mathbb{P}_{\text{L}}^{s,e} 
=& \prod_{l=0}^{f(r)} \left[1- \exp \left(-\frac{\left(H-(l+1/2)H/(f(r)+1) \right )^2}{2\sigma^2} \right ) \right ]\\
\approx & 
\frac{1}{1+ C \exp\left(-B\left(\theta_{s,e} - C\right)\right)}
\end{split}
\end{equation}
where $f(r) = \lfloor r \sqrt{\rho_1 \rho_2 }-1\rfloor$, $r= l_{t,e}$ is the ground distance between $s$ and $e$, $\rho_1$ is the ratio of built-up land area to the total land area, $\rho_2$ is the mean number of buildings per
unit area (buildings/km$^2$), and $\sigma$ is the scale parameter of the Rayleigh probability density function (PDF), which gives the height distribution of buildings in meters. The environment parameters ($\rho_1, \rho_2, \sigma$) for typical environments are suburban ($0.1, 750, 8$), urban ($0.3, 500, 15$), dense urban ($0.5, 300, 20$), and highrise urban ($0.5, 300, 50$), respectively. Note that increasing $H$ will increase the smoothness of the plot of the product, and $\mathbb{P}_{\text{L}}^{s,e} $ can be considered as a continuous function of elevation angle and the environment parameters for a large $H$ \cite{al2014optimal}. The approximation to the LoS probability of an ATG link is given by a sigmoid function, where $C$ and $B$ are constant values depending on the aforementioned environment and $\theta_{s,e}$ is the elevation angle in radians, which is given by $\theta_{s,e} = {\tan^{-1}} \left(\frac{H}{l_{s,e}}\right) $. The ATG channel model also applies to the received jamming power $P_{j,d}$ with LoS and NLoS probability given by $\mathbb{P}_{\text{L}}^{j,d}$ and $\mathbb{P}_{\text{N}}^{j,d}$.

\begin{figure}[t]
\centering
     \includegraphics[width=0.481\textwidth]{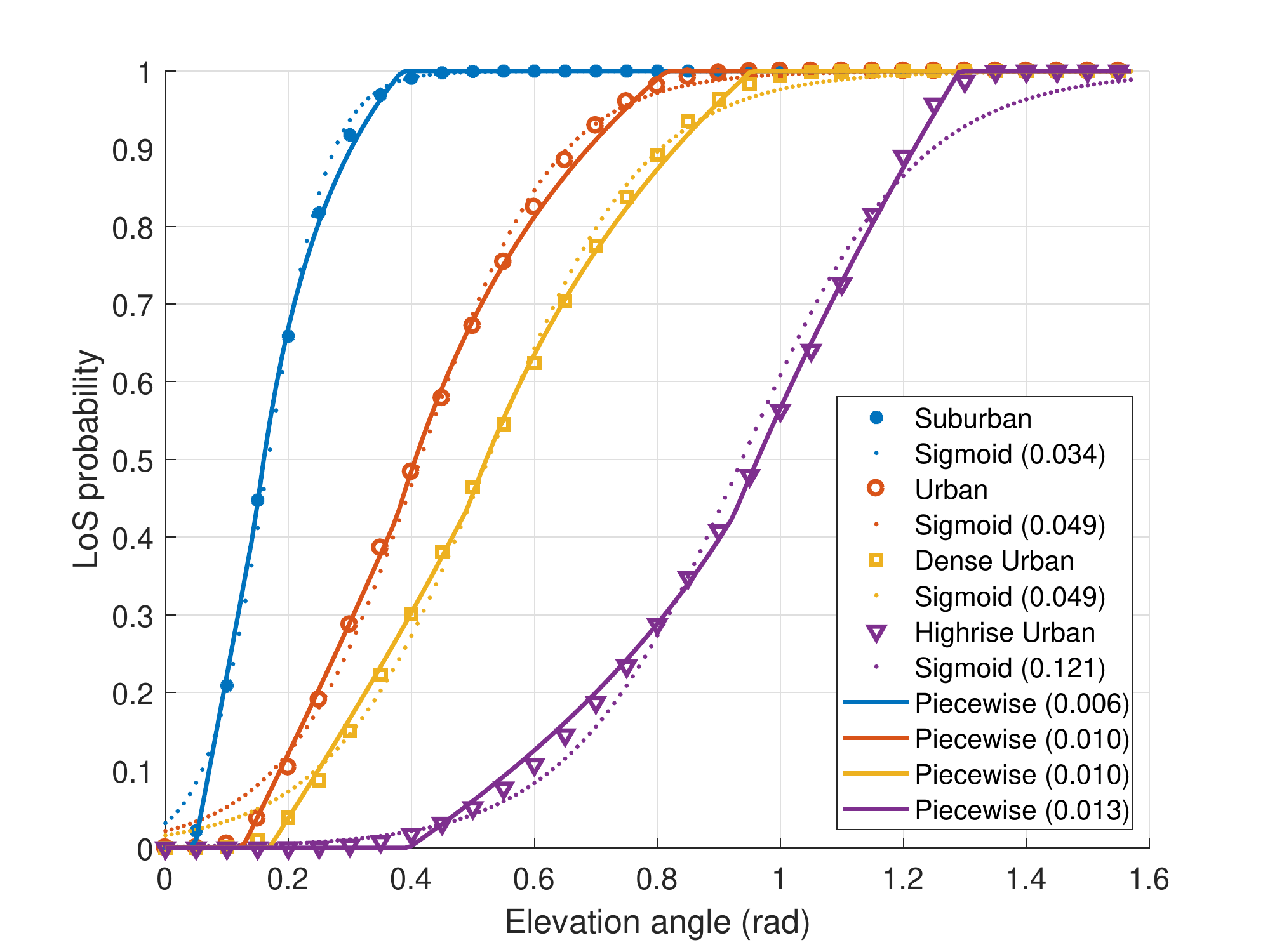}
  \caption{Plots of the LoS probabilities for different urban environments based on the exact product, the sigmoid fitting and the proposed piecewise fitting. The goodness of fit measure by root-mean-square error is given inside the brackets.}
 \label{fig:net_model}
\end{figure}

\paragraph{Proposed piecewise fitting}
Although the sigmoid fitting provides a reasonably good approximation to the actual LoS probability, it provides a non-linear relationship to $l_{s,e}$ and has been shown to only give analytical forms which are non-trivial to evaluate. Based on the value of the elevation angle, each curve of the LoS probability can be divided into three approximated regions: LoS region for the link in a pure LoS state with high elevation angle, NLoS region for the the link in a pure NLoS state with low elevation angle, and a transitional region where the link is in mixed LoS/NLoS states. The transitional region can be further divided into two regions based on the slope of the curve as the elevation angle increases. As a result, we propose a piecewise fitting model in \eqref{eq_fitting_3}, which not only gives a meaningful relationship between the LoS probability and $l_{s,e}$ but also provides a tractable way to calculate the SCP as we will show later. The piecewise approximation is given by 
\begin{equation} \label{eq_fitting_3}
\mathbb{P}_{\text{L}}^{s,e}=
\begin{cases}
1, &\text{if } \theta_{s,e} > \tan^{-1} \frac{H}{{\ell_1}} \\
c_3\cot(\theta_{s,e}) +c_4, &\text{if }  \tan^{-1} \frac{H}{{\ell_2}}< \theta_{s,e} \leq \tan^{-1} \frac{H}{{\ell_1}} \\
0, &\text{if } \theta_{s,e} \leq \tan^{-1} \frac{H}{{\ell_3}}  \\
c_1\tan(\theta_{s,e}) +c_2, &\text{otherwise}\\
\end{cases}
\end{equation}
where ${\ell_1} = H\frac{1-c_4}{c_3}$, ${\ell_2} = \frac{2Hc_1}{c_4-c_2-\sqrt{(c_4-c_2)^2+4c_1c_3} } $, ${\ell_3} = -H\frac{c_1}{c_2}$, $c_1$, $c_2$, $c_3$ and $c_4$ are given in Table \ref{tb_fitting3} and obtained by Algorithm \ref{UAV_algorithm_1}, and the corresponding results for different typical environments are plotted in Fig. \ref{fig:net_model}.

\begin{algorithm}[t] 
\SetKwInput{KwData}{\bf{Given}} 

\KwData{Environment parameters ($\rho_1, \rho_2, \sigma$), thresholds: $c_l = 0.005$, $c_m = 0.5$, and $c_u = 0.995$.}

Generate $10^4$ evenly spaced samples from $[0, \pi/2]$, and obtain the corresponding samples of LoS probabilities with \eqref{eq_LoS_NLoS}. \\

Create a new dataset $D_{t_1}$ from the samples of LoS probabilities whose values are bigger than $c_l$ and smaller than $c_m$.\\

Find the coefficients $c_1$ and $c_2$ of the model function $c_1\tan(\theta_{s,e}) +c_2$ by fitting the function to $D_{t_1}$ with the trust region reflective least squares algorithm.\\

Let $c_1\tan(\theta_{s,e}) +c_2=0$, the maximum elevation angle for having only the NLoS link is $\tan^{-1} \left(-\frac{c_2}{c_1} \right)$. \\

Create a new dataset $D_{t_2}$ from the samples of LoS probabilities whose values are bigger than $c_m$ and smaller than $c_u$.\\

Find the coefficients $c_3$ and $c_4$ of the model function $c_3\cot(\theta_{s,e}) +c_4$ by fitting the function to $D_{t_2}$ with the trust region reflective least squares algorithm.\\

Let $c_3\cot(\theta_{s,e}) +c_4=1$, the minimum elevation angle for having only the LoS link is $\tan^{-1} \left(\frac{c_3}{1-c_4} \right)$. \\

The elevation angle at the inflection point, where the LoS probability changes from being $c_1\tan(\theta_{s,e}) +c_2$ to $c_3\cot(\theta_{s,e}) +c_4$ is $\tan^{-1} \left(\frac{c_4-c_2-\sqrt{(c_4-c_2)^2+4c_1c_3} }{2c_1} \right)$, is obtained by solving $c_1\tan(\theta_{s,e}) +c_2=c_3\cot(\theta_{s,e}) +c_4$, and then selecting the root which provides a better fit to the samples of LoS probabilities. \\

\caption{Finding parameters $c_1, c_2, c_3 \text{ and } c_4$ for link $(s,e)$ given environment parameters ($\rho_1, \rho_2, \sigma$). }
 \label{UAV_algorithm_1}
\end{algorithm}

\begin{table}[t]
\caption{The parameters for \eqref{eq_fitting_3} and \eqref{eq_fitting_32}.}
\label{tb_fitting3}
\begin{center}
\begin{tabular}{ |c|c|c|c|c| } 
\hline
&Suburban & Urban & Dense urban & Highrise urban \\
\hline
$c_1$& 4.215&1.581&1.201&0.4717\\
$c_2$&-0.2007&-0.1991&-0.2051&-0.1972\\
$c_3$&-0.1341&-0.3618&-0.4864&-1.223\\
$c_4$&1.331&1.341&1.346&1.351\\
\hline
\end{tabular}
\end{center}
\end{table}

Since $\tan(\theta_{s,e}) = \frac{H}{l_{t,e}}$, \eqref{eq_fitting_3} can be further written as
\begin{equation} \label{eq_fitting_32}
\mathbb{P}_{\text{L}}^{s,e}=
\begin{cases}
1, &\text{if } l_{t,e} < {\ell_1}\\
c_3l_{t,e}/H +c_4, &\text{if } {\ell_1}\leq l_{t,e} < {\ell_2}\\
c_1 H/l_{t,e} +c_2, &\text{if } {\ell_2}\leq l_{t,e} < {\ell_3}\\
0, &\text{otherwise}.
\end{cases}
\end{equation}

In this figure, the proposed piecewise fitting outperforms the sigmoid fitting in the goodness of fit measured by root-mean-square error. The former is at least four times better than the latter in the suburban environment and nine times better in the highrise urban environment. 
The transition region is further approximated by a linear function of the variable $l_{t,e}$ and another linear function of the reciprocal of $l_{t,e}$. In the transition region, $\mathbb{P}_{\text{L}}^{s,e}$ increases as $l_{t,e}$ decreases. 
%


\subsubsection{Ground-to-ground channel model}
For ground communications, the log-distance path loss model is used to characterize the legitimate link from $s$ to ${d}$. The received power at $d$ from $s$ is written as
\begin{equation}
P_{{s,d}}=P_{s}(\lambda/4\pi)^2 l_{{s},{d}}^{-\beta}|h_{{s,d}}|^2
\end{equation}
where $\lambda$ is the carrier wavelength, $|h_{{s,d}}|^2$ is the channel gain associated with the Rayleigh fading, and $\beta$ is the path loss exponent for ground communications. 

\section{Derivation of the SCP}
\subsection{Secrecy capacity}
Let $\Phi =\{1,2,...,n\}$ denote the collection of $n$ UEDs. The secrecy capacity of a link is the difference between the capacity of the main link and the capacity achieved via a collection of wire-tap links. The general form is given as \cite{chen2015physical}  
\begin{equation}\label{eq_secrecy_capacity}
\begin{split}
C_s
& = \left [\log_2\left({\frac{1+\frac{P_{s,d}}{P_{j,d}+N_0}}{ 1+\max_{{e} \in \Phi} \left(\frac{P_{s,e}}{P_{j,e}+N_e} \right)}}\right) \right]^+\\
&\geq \left [\log_2\left({\frac{\Gamma_1}{ \max_{{e} \in \Phi}\Gamma_2}}\right) \right]^+\\
\end{split}
\end{equation}
where $N_0$ is the additive white Gaussian noise (AWGN) power at the ground receiver $d$, $N_e$ is the AWGN power at a UED, $[x]^+$ denotes $\max(0,x)$ and the approximation holds over interference-limited channels \cite{chen2016dual}. '$\geq$' holds if the receivers on the UEDs have the low-noise figure by using the thermoelectric cooling and new materials \cite{blank1998exploring, alford2005thermoelectric,weinreb2007design}.
$\Gamma_1 = 1+\frac{P_{s,d}}{P_{j,d}+N_0}$ represents one plus the signal-to-interference-plus-noise ratio ($1+\text{SINR}$) from $s$ to $d$, and $\Gamma_2 = { 1+\frac{P_{s,e}}{P_{j,e}} }$ represents $1 + \text{SIR}$ from $s$ to $e$, where the interference is due to UAV jammer $j$.
 
\subsection{SCP formulation}
The SCP is defined as
\begin{equation}\label{security_con}
\begin{split}
p_{c}&=\mathbb{P} {\left (C_s > \mathcal{R}_t \right)}
\geq\mathbb{P} {\left ({\frac{\Gamma_1}{ \max_{{e} \in \Phi}\Gamma_2}} > 2^{\mathcal{R}_t} \right)}\\
\end{split}
\end{equation}
where $\mathcal{R}_t \geq 0$ is the target secrecy rate. Besides, $p_{c}=\mathbb{P} {\left (C_s > \mathcal{R}_t \right)} \approx \mathbb{P} {\left ({\frac{\Gamma_1}{ \max_{{e} \in \Phi}\Gamma_2}} > 2^{\mathcal{R}_t} \right)}$, when both $d$ and UEDs are working in an interference-limited environment as the result of a large jamming signal. To evaluate the above inequality, we require the following calculations.

Since $\Gamma_1$ is a random variable related to $|h|^2$, the CDF of $\Gamma_1$ can be written as
\begin{equation}
\begin{split}
F&_{\Gamma_1}({\gamma_1}) = \mathbb{P} \left( 1+\frac{P_{s,d}}{P_{j,d}+N_0} <{\gamma_1} \right)\\
=&\mathbb{P} \left( \frac{|h|^2}{\Lambda(\eta_{\text{L}}) } <  {\gamma_1}-1\right)\mathbb{P}_{\text{L}}^{j,d} + \mathbb{P} \left( \frac{|h|^2}{\Lambda(\eta_{\text{N}})} <  {\gamma_1}-1 \right)\mathbb{P}_{\text{N}}^{j,d}\\
= &1- \exp \left ( (1-{\gamma_1})\Lambda(\eta_{\text{L}})\right)\mathbb{P}_{\text{L}}^{j,d} - \exp \left ((1-{\gamma_1}) \Lambda(\eta_{\text{N}})\right)\mathbb{P}_{\text{N}}^{j,d}\\
\end{split}
\end{equation}
where $\Lambda(\eta) = \frac{l_{s,d}^ \beta}{P_s} \left(\eta l_{j,d}^{-\alpha} {P_j} +N_0\right)$, $l_{j,d}^2 = l_{t,j}^2 + l_{s,d}^2 - 2l_{t,j} l_{s,d} \cos\left( \varphi_{j'} - \varphi_{d}\right) + H^2$, $j'$ denotes the projection of $j$ on the xy-plane. $\varphi_{j'}$ and $\varphi_{d}$ denote the angles of $l_{s,j'}$ and $l_{s,d}$ measured counterclockwise from the x-axis. Thus, the corresponding PDF is given by $f_{\Gamma_1}({\gamma_1}) = {\frac {{\rm{d}}}{{\rm{d}}\,{\gamma_1}}} F_{\Gamma_1}({\gamma_1})$ such that 
\begin{equation}
\begin{split}
f_{\Gamma_1}({\gamma_1}) 
= &\Lambda(\eta_{\text{L}})\exp  \left((1-{\gamma_1}) \Lambda(\eta_{\text{L}}) \right )\mathbb{P}_{\text{L}}^{j,d}+ \AND \Lambda(\eta_{\text{N}})\exp  \left((1-{\gamma_1}) \Lambda(\eta_{\text{N}}) \right )\mathbb{P}_{\text{N}}^{j,d}.\\
\end{split}
\end{equation}

Let $e'$ denote the projection of the node $e$ on the xy-plane, and $\phi = \varphi_{j'} - \varphi_{e'}$. $\Gamma_2$ is a random variable  related to the position of one UED $e$, which is uniformly distributed on the disk $O(t,R_{1})$. Thus, the CDF of $\Gamma_2$ is given by
\begin{equation} \label{func_1}
\begin{split}
F_{\Gamma_2}({\gamma_2}) &= \mathbb{E}_e\left[F_1({\gamma_2}, l_{t,e},\phi)  \right]\\
&= \frac{1}{\pi R_{1}^2} \int_0^{2\pi} \int_0^{R_{1}}  F_1({\gamma_2}, l_{t,e},\phi) l_{t,e} \;  {\rm{d}} l_{t,e} {\rm{d}} \phi\\
\end{split}
\end{equation}
where 
\begin{equation}
F_1({\gamma_2},l_{t,e},\phi) ={g}({\gamma_2}, \eta_{\text{L}},l_{t,e},\phi)\mathbb{P}_{\text{L}}^{s,e} + {g}({\gamma_2}, \eta_{\text{N}},l_{t,e},\phi)\mathbb{P}_{\text{N}}^{s,e}
\end{equation}
and
$
{g} ({\gamma},\eta,\ell,\phi) = \mathbbm{1} \left(1+\frac{(H^2+\ell^2)^\frac{-\alpha}{2}\eta P_s}{\left(l_{t,j}^2+\ell^2-2 l_{t,j} \ell\cos \phi\right)^\frac{-\alpha}{2}P_j} \leq {\gamma}\right).
$

\subsection{Calculation of the SCP}
 
When $\mathbb{P}_{\text{L}}^{s,e}$ is given by \eqref{eq_LoS_NLoS}, it is hard to find a closed-form expression for \eqref{func_1}. To make progress, we use \eqref{eq_fitting_3} instead, which can yield a closed-form expression for \eqref{func_1} with any given $R_{1}$ and $H$. With the help of \eqref{eq_fitting_3}, we can divide the LoS probability into multiple regions, and calculate their contributions to $F_{\Gamma_2}({\gamma_2})$ independently. This decomposition technique to obtain \eqref{func_1} seems to be tedious, but necessary, because it leads to a tractable solution. Throughout the following discussion, we refer to a number of propositions, which will be presented in Section \ref{sec_props}.

 
For $R_{1} < {\ell_1}$, where $l_1$ is considered in \eqref{eq_fitting_32}, the ATG link between transmitter $s$ and UED $e$ is always in a pure LoS state since $l_{t,e}\leq R_{1}$. Following Proposition \ref{th_UAV_1}, we have
\begin{equation}
\begin{split}
F_{\Gamma_2}({\gamma_2}) 
&=\int_0^{2\pi} \int_0^{R_{1}}  {g}({\gamma_2}, \eta_{\text{L}},l_{t,e},\phi)\frac{l_{t,e} }{\pi R_{1}^2}  \;  {\rm{d}} l_{t,e} {\rm{d}} \phi = F_{\Gamma_3}({\gamma_2}, \eta_{\text{L}},R_{1}).
\end{split}
\end{equation}

For ${\ell_1} \leq R_{1} < {\ell_2}$, the ATG link between transmitter $s$ and UED $e$ is in a pure LoS state if $l_{t,e}< {\ell_1}$, and the ATG link between transmitter $s$ and UED $e$ is in the mixed LoS/NLoS states with $\mathbb{P}_{\text{L}}^{s,e}=c_3 l_{t,e}/H +c_4$ if ${\ell_1}\leq l_{t,e} \leq R_{1}$.  Following Propositions \ref{th_UAV_1} and \ref{th_UAV_3}, we have
\begin{equation}
\begin{split}
F_{\Gamma_2}({\gamma_2}) 
&=\int_0^{2\pi} \int_0^{{{\ell_1} }}  {g}({\gamma_2}, \eta_{\text{L}},l_{t,e},\phi)\frac{l_{t,e} }{\pi R_{1}^2}  \;  {\rm{d}} l_{t,e} {\rm{d}} \phi \, +  \AND \int_0^{2\pi} \int_{{\ell_1} }^{R_{1}} F_1({\gamma_2}, l_{t,e},\phi)  \frac{l_{t,e}}{\pi R_{1}^2}  \;  {\rm{d}} l_{t,e} {\rm{d}} \phi \\
&= F_{\Gamma_3}({\gamma_2}, \eta_{\text{L}},{{\ell_1} }) +F_{\Gamma_5}({\gamma_2}, \eta_{\text{L}}, \eta_{\text{N}},R_{1}) -F_{\Gamma_5}({\gamma_2},  \eta_{\text{L}}, \eta_{\text{N}}, {\ell_1}).
\end{split}
\end{equation}

For ${\ell_2}\leq R_{1} < {\ell_3}$, the ATG link between transmitter $s$ and UED $e$ is in a pure LoS state if $l_{t,e}<{\ell_1}$, the ATG link between transmitter $s$ and UED $e$ is in the mixed LoS/NLoS states with $\mathbb{P}_{\text{L}}^{s,e}=c_3 l_{t,e}/H +c_4$ if ${\ell_1}\leq l_{t,e} < {\ell_2}$,  and the ATG link between transmitter $s$ and UED $e$ is in the mixed LoS/NLoS states with $\mathbb{P}_{\text{L}}^{s,e}=c_1 H/l_{t,e} +c_2$ if ${\ell_2}\leq l_{t,e} \leq R_{1}$. Following Propositions \ref{th_UAV_1}, \ref{th_UAV_2} and \ref{th_UAV_3}, we have
\begin{equation}
\begin{split}
F&_{\Gamma_2}({\gamma_2}) 
=\int_0^{2\pi} \int_0^{{{\ell_1} }}  {g}({\gamma_2}, \eta_{\text{L}},l_{t,e},\phi)\frac{l_{t,e} }{\pi R_{1}^2}  \;  {\rm{d}} l_{t,e} {\rm{d}} \phi \, +\AND \int_0^{2\pi} \int_{{\ell_1} }^{{R_1}} F_1({\gamma_2}, l_{t,e},\phi)  \frac{l_{t,e}}{\pi R_{1}^2}  \;  {\rm{d}} l_{t,e} {\rm{d}} \phi \,\\
=& F_{\Gamma_3}({\gamma_2}, \eta_{\text{L}},{{\ell_1} }) +F_{\Gamma_5}({\gamma_2}, \eta_{\text{L}}, \eta_{\text{N}},{\ell_2}) - F_{\Gamma_5}({\gamma_2},  \eta_{\text{L}}, \eta_{\text{N}}, {\ell_1}) 
+\AND  F_{\Gamma_4}({\gamma_2}, \eta_{\text{L}}, \eta_{\text{N}}, R_{1}) -F_{\Gamma_4}({\gamma_2},  \eta_{\text{L}}, \eta_{\text{N}}, {\ell_2}).
\end{split}
\end{equation}

For $R_{1} \geq {\ell_3}$, the ATG link between transmitter $s$ and UED $e$ is in a pure LoS state if $l_{t,e}< {\ell_1}$, the ATG link between transmitter $s$ and UED $e$ is in the mixed LoS/NLoS states with $\mathbb{P}_{\text{L}}^{s,e}=c_3 l_{t,e}/H +c_4$ if ${\ell_1}\leq l_{t,e} < {\ell_2}$, the ATG link between transmitter $s$ and UED $e$ is in the mixed LoS/NLoS states with $\mathbb{P}_{\text{L}}^{s,e}=c_1 H/l_{t,e} +c_2$ if ${\ell_2}\leq l_{t,e} < {\ell_3}$, and the ATG link between transmitter $s$ and UED $e$ is in a pure NLoS state if $ {\ell_3} \leq l_{t,e} \leq R_{1}$. Following Propositions \ref{th_UAV_1}, \ref{th_UAV_2} and \ref{th_UAV_3}, we have
\begin{equation}
\begin{split}
F_{\Gamma_2}({\gamma_2}) 
=&\int_0^{2\pi} \int_0^{{{\ell_1} }}  {g}({\gamma_2}, \eta_{\text{L}},l_{t,e},\phi)\frac{l_{t,e} }{\pi R_{1}^2}  \;  {\rm{d}} l_{t,e} {\rm{d}} \phi\, +\AND  \int_0^{2\pi} \int_{{\ell_1} }^{{\ell_3}} F_1({\gamma_2}, l_{t,e},\phi)  \frac{l_{t,e}}{\pi R_{1}^2}  \;  {\rm{d}} l_{t,e} {\rm{d}} \phi \,+ \\&  \int_0^{2\pi} \int_{{\ell_3} }^{R_{1}}  {g}({\gamma_2}, \eta_{\text{N}},l_{t,e},\phi)\frac{l_{t,e} }{\pi R_{1}^2}  \;  {\rm{d}} l_{t,e} {\rm{d}} \phi\\
= & F_{\Gamma_3}({\gamma_2}, \eta_{\text{L}},{{\ell_1} }) +F_{\Gamma_5}({\gamma_2}, \eta_{\text{L}}, \eta_{\text{N}},{\ell_2}) -F_{\Gamma_5}({\gamma_2},  \eta_{\text{L}}, \eta_{\text{N}}, {\ell_1}) 
+\\& F_{\Gamma_4}({\gamma_2}, \eta_{\text{L}}, \eta_{\text{N}},{\ell_3}) - F_{\Gamma_4}({\gamma_2},  \eta_{\text{L}}, \eta_{\text{N}}, {\ell_2}) 
+ F_{\Gamma_3}({\gamma_2}, \eta_{\text{N}}, R_{1}) - \AND  F_{\Gamma_3}({\gamma_2}, \eta_{\text{N}},{\ell_3} ).
\end{split}
\end{equation}

As a result of the above decomposition, $p_c$ is lower bounded by
\begin{equation}\label{SSP_lb}
\begin{split}
\mathbb{P} \left(\frac{\Gamma_1}{\max_{{e} \in \Phi}{\Gamma_2}}>2^{\mathcal{R}_t} \right) = \int_1^\infty  F_{\Gamma_2}\left(\frac{\gamma_1}{2^{\mathcal{R}_t}}\right)^{n} f_{\Gamma_1} (\gamma_1) \, {\rm{d}} \gamma_1.\\
\end{split}
\end{equation}

\subsection{Propositions} \label{sec_props}
All three propositions given below rely on Lemma \ref{lm_UAV_2}, which is given in the Appendix. 
\begin{table}[t]
\centering
\caption{Table of conditions.}
\label{condition_tabel}
\begin{tabular}{ll}
\hline
Case index                           & Condition                \\ \hline
1 & ${A}<0, B \leq 0,{r}^2{A}<B, \, \frac{-{r}^2{A} - B}{2 {r}} \leq l_{t,j},l_{t,j}\neq \sqrt{B{A}}$
\\
2 & ${A}<0, B \leq 0,{r}^2{A}=B, \, \frac{-{r}^2{A} - B}{2 {r}} \leq l_{t,j},l_{t,j}\neq \sqrt{B{A}}$
 \\ 
3 & ${A}<0, B \leq 0, {r}^2{A}<B, \sqrt{B{A}} < l_{t,j} < \frac{-{r}^2{A} - B}{2 {r}}$\\
4 & ${A}<0, B \leq 0, {r}^2{A}>B, \frac{-{r}^2{A} - B}{2 {r}} \leq l_{t,j} $\\
5 & Other cases under ${A}<0, B \leq 0$\\
6& ${A}>0, B \geq 0, {r}^2{A}>B, \frac{{r}^2{A} + B}{2 {r}} \leq l_{t,j},l_{t,j}\neq \sqrt{B{A}} $\\
7& ${A}>0, B \geq 0, {r}^2{A}=B, \frac{{r}^2{A} + B}{2 {r}} \leq l_{t,j},l_{t,j}\neq \sqrt{B{A}}$\\
8& ${A}>0, B \geq 0, {r}^2{A}>B, \sqrt{B{A}} < l_{t,j} < \frac{{r}^2{A} + B}{2 {r}}$\\
9& ${A}>0, B \geq 0, {r}^2{A}<B, \frac{{r}^2{A} + B}{2 {r}} \leq l_{t,j}$\\
10 & ${A}<0, B > 0, l_{t,j}\geq  \frac{\left |{r}^2{A} + B\right|}{2 {r}} $\\
11 & ${A}<0, B > 0, -{r}^2{A}>B, l_{t,j} < \frac{-{r}^2{A} - B}{2 {r}}$\\
12& Other cases under ${A}<0, B > 0$\\
13& ${A}>0, B < 0, -{r}^2{A}>B, l_{t,j}< \frac{-{r}^2{A} - B}{2 {r}} $\\
14& ${A}>0, B < 0,l_{t,j}\geq \frac{|{r}^2{A} + B |}{2 {r}} $\\
15& ${A}>0, B < 0,-{r}^2{A}<B, l_{t,j}< \frac{{r}^2{A} + B}{2 {r}} $\\
16 & ${A}=0, \frac{ l_{t,j}^2-H^2 }{2  {r}} \leq l_{t,j},  l_{t,j}> H$\\
17 & ${A}=0, \frac{H^2- l_{t,j}^2 }{2  {r}}   \leq l_{t,j},  l_{t,j}< H$\\
18 & ${A}=0,  l_{t,j} = H$\\
19 & ${A}=0, l_{t,j} < \frac{H^2-l_{t,j}^2 }{2 {r} }$\\
\hline
\end{tabular}
\end{table}
\begin{prop}\label{th_UAV_1}
{\rm{Let ${r} \leq R_{1}$, assuming that the link $(s,e)$ is always in the same channel state (i.e, pure LoS or pure NLoS) with gain $\eta$, the CDF of $\Gamma_2$ conditioning on that $e$ is only active for $l_{t,e}\leq {r}$ is given by
\begin{equation}\SMALL \label{eq_4}
\begin{split}
F_{\Gamma_3}(y, \eta,{r}) &= \int_0^{2\pi} \int_0^{{r}}  \frac{{g}({\gamma_2}, \eta,l_{t,e},\phi)l_{t,e} }{\pi R_{1}^2}  \;  {\rm{d}} l_{t,e} {\rm{d}} \phi
 = \frac{S_1(y,\eta, {r})}{\pi R_{1}^2/2}
\end{split}
\end{equation}
where }}{ 
\begin{equation}\label{eq_s1_cases}
S_1(y,\eta, {r}) =
\begin{cases}
D_1 + D_2 -\frac{1}{2}D_3, & \text{ \rm{cases  1, 9, 10} } \\
D_2 +D_4, & \text{ \rm{cases 2, 7} } \\
\frac{1}{2}\pi {r}^2 - D_3, & \text{ \rm{cases 3, 11} }  \\
D_1 +D_2 +\frac{1}{2}D_3, & \text{ \rm{cases 4, 6, 14} } \\
 \frac{1}{2}\pi {r}^2,  &\text{ \rm{cases 5, 12, 13, 19} }\\
D_3, & \text{ \rm{cases 8, 15} } \\
D_5, & \text{ \rm{cases 16, 17, 18} } \\
0, & \text{ \rm {otherwise}}
\end{cases}
\end{equation}
{\rm and all the cases with index numbers are given in Table \ref{condition_tabel},  $D_1= \frac{{A} B-l_{t,j}^2}{2 {A}^2} \cot ^{-1}\left(\frac{{a_1}-2 B}{\sqrt{4 l_{t,j}^2 {r}^2-{a_1}^2}}\right)$,  ${A} = 1-\left((y-1)\frac{P_j}{\eta P_s}\right)^\frac{2}{\alpha}$, $B= l_{t,j}^2- (1-{A}) H^2$, ${a_1} ={A} {r}^2+B$,
$D_2= \frac{\pi {r}^2}{4} - \frac{{A}}{4 {A}^2} \sqrt{4 l_{t,j}^2 {r}^2-{a_1}^2} - \frac{{a_1}{A}-l_{t,j}^2}{2 {A}^2} \csc ^{-1}\left(\frac{2 l_{t,j} {r}}{{a_1}}\right)$,
$D_3= \frac{  l_{t,j}^2-{A} B}{2 {A}^2} \pi$,
$D_4=\frac{({a_1}-2 B) \sqrt{4 j^2 {r}^2-{a_1}^2}}{4 {A}^2 {r}^2}$,
$D_5=\frac{1}{2} {r}^2 \sec ^{-1}\left(\frac{2 l_{t,j} {r}}{{a_1}}\right)-\frac{\left(l_{t,j}^2-H^2\right) \sqrt{4 l_{t,j}^2 {r}^2 - {\left(H^2-l_{t,j}^2\right)^2} }}{8 l_{t,j}^2 }$.  }

}
\end{prop}

\begin{IEEEproof}
\begin{equation}
\begin{split}
&\frac{1}{\pi R_{1}^2}\int_0^{2\pi} \int_0^{{r}}  {g}(y, \eta_{\text{L}},l_{t,e},\phi){l_{t,e} } \;  {\rm{d}} l_{t,e} {\rm{d}} \phi
\\& =\frac{1}{\pi R_{1}^2} \int_{o_\phi(\eta)} \int_{o_{l_{t,e}}(\eta)}  {l_{t,e} } \;  {\rm{d}} l_{t,e} {\rm{d}} \phi
\\& =\frac{{r}^2}{R_{1}^2}- \frac{1}{\pi R_{1}^2} \int_{{o'}_\phi(\eta)} \int_{{o'}_{l_{t,e}}(\eta)}  {l_{t,e} } \;  {\rm{d}} l_{t,e} {\rm{d}} \phi
\end{split}
\end{equation}
where ${o_\phi}(\eta)$ and $o_{l_{t,e}}(\eta)$ are the domains so that $1+\frac{(H^2+l_{t,e}^2)^\frac{-\alpha}{2}\eta P_s}{\left(l_{t,j}^2+l_{t,e}^2-2 l_{t,j} l_{t,e}\cos \phi\right)^\frac{-\alpha}{2}P_j} \leq y$, while ${{o'}_\phi}(\eta)$ and ${o'}_{l_{t,e}}(\eta)$ are the domains so that $1+\frac{(H^2+l_{t,e}^2)^\frac{-\alpha}{2}\eta P_s}{\left(l_{t,j}^2+l_{t,e}^2-2 l_{t,j} l_{t,e}\cos \phi\right)^\frac{-\alpha}{2}P_j} > y$.
Given Lemma \ref{lm_UAV_2}, \eqref{eq_4} is achieved. Thus, we conclude the proof. 
\end{IEEEproof}

\begin{prop} \label{th_UAV_2}
{\rm{Let $r\leq R_{1}$, assuming that the mixed LoS/NLoS states exist for link $(s,e)$, and the (quasi-) LoS probability is $\mathbb{P}_{\text{L}}^{s,e}=c_1 H/l_{t,e} +c_2$, the CDF of $\Gamma_2$ conditioning on the UED $e$ being active within the disk $O(t,r)$ is given by
\begin{equation} \label{eq_th2_1}
\begin{split}
F_{\Gamma_4}(y, \eta_{\text{L}}, \eta_{\text{N}},r) =& \frac{2}{\pi R_{1}^2}\big( c_2 S_1(y,\eta_{\text{L}}, r)+ c_1H S_2(y,\eta_{\text{L}}, r)+ \AND (1-c_2) S_1(y,\eta_{\text{N}}, r)-c_1 H S_2(y,\eta_{\text{N}}, r)\big)
\end{split}
\end{equation}
\begin{equation}
S_2(y,\eta, r) =
\begin{cases}
G_1 - G_2 +G_3 -G_4 , & \text{ \rm{cases 1, 2} } \\
\pi  r + G_1- 2G_2, & \text{ \rm{cases 3} }  \\
G_3+ G_4 -G_2, & \text{ \rm{cases 4, 14} } \\
\pi r,  &\text{ \rm{cases 5, 12, 13, 19} } \\
G_3 +G_4 -G_5 , & \text{ \rm{cases 6, 7} } \\
-G_5, & \text{ \rm{case 8} } \\
G_3- G_4, & \text{ \rm{cases 9, 10} } \\
\pi  r- G_2, & \text{ \rm{case 11} }  \\
-G_2, & \text{ \rm{case 15} }   \\
G_6, & \text{ \rm{cases 16, 17} } \\
\frac{1}{2}\pi r, & \text{ \rm{case 18} }\\
0, & \text{ \rm {otherwise}}
\end{cases}
\end{equation}
and all the cases with index are given in Table \ref{condition_tabel},
$
G_0=\frac{2 \sqrt{-{A} B+l_{t,j}^2} }{-{A}},
$
$
G_1=G_0 \mathsf{E}\left(\sec ^{-1}\left(\frac{-l_{t,j}}{\sqrt{B{A}}}\right)|\frac{l_{t,j}^2}{l_{t,j}^2-{A} B}\right),
$
$
G_2=G_0 \mathsf{E}\left(\frac{l_{t,j}^2}{l_{t,j}^2-{A} B}\right),
$
$
G_3=\frac{\sqrt{4 l_{t,j}^2 r^2-{a_1}^2}-2 {A} {r}^2 \sec ^{-1}\left(\frac{2 l_{t,j} r}{{a_1}}\right)}{-2{A}r},
$
$
G_4={G_0} \mathsf{E}\left(\sec ^{-1}\left(\frac{2 l_{t,j} r}{{a_1}}\right)|\frac{l_{t,j}^2}{l_{t,j}^2-{A} B}\right)/{2},
$
$
G_5 =G_0 \mathsf{E}\left(\sec ^{-1}\left(\frac{l_{t,j}}{\sqrt{B{A}}}\right)|\frac{l_{t,j}^2}{l_{t,j}^2-{A} B}\right),
$
$
G_6= r \cos ^{-1}\left(\frac{l_{t,j}^2-H^2}{2 l_{t,j} r}\right) -\frac{l_{t,j}^2-H^2}{2 l_{t,j}} {  \ln \frac{  \sqrt{4 l_{t,j}^2 r^2- {\left(H^2-l_{t,j}^2\right)^2}}+2l_{t,j}r } {\left |l_{t,j}^2-H^2\right |}  }
$, $\mathsf{E}(\cdot)$ and $\mathsf{E}(\cdot \mid \cdot)$ denote complete and incomplete elliptic integral of the 2nd kind \cite{NIST:DLMF}. 
}}
\end{prop}

\begin{IEEEproof}
\begin{equation}\label{pro2_eq}
\begin{split}
\int_0^{2\pi} &\int_0^{{r}}  ({g}(y, \eta_{\text{L}},l_{t,e},\phi)\mathbb{P}_{\text{L}}^{s,e} + {g}(y, \eta_{\text{N}},l_{t,e},\phi)\mathbb{P}_{\text{N}}^{s,e} ){l_{t,e} } \;  {\rm{d}} l_{t,e} {\rm{d}} \phi\\
=&\int_{o_{\phi}(\eta_\text{L})} \int_{o_{l_{t,e}}(\eta_\text{L})} \left(c_1 H+ c_2{l_{t,e} } \right)\;  {\rm{d}} l_{t,e} {\rm{d}} \phi +\AND \int_{o_{\phi}(\eta_\text{N})} \int_{o_{l_{t,e}}(\eta_\text{N})} \left(-c_1 H+ (1-c_2){l_{t,e} } \right) \;  {\rm{d}} l_{t,e} {\rm{d}} \phi \\
 = &{c_1 H }\int_{o_{\phi}(\eta_\text{L})} \int_{o_{l_{t,e}}(\eta_\text{L})} \;  {\rm{d}} l_{t,e} {\rm{d}} \phi + 2 {c_2 S_1(y,\eta_{\text{L}}, {r})}- \AND {c_1 H }\int_{o_{\phi}(\eta_\text{N})} \int_{o_{l_{t,e}}(\eta_\text{N})} \;  {\rm{d}} l_{t,e} {\rm{d}} \phi + 2{(1-c_2)S_1(y,\eta_{\text{N}}, {r})} \\
\end{split}
\end{equation}
where 
$
\int_{o_{\phi}(\eta)} \int_{o_{l_{t,e}}(\eta)}  \;  {\rm{d}} l_{t,e} {\rm{d}} \phi  = {2 \pi {r}} - \int_{{o'}_{\phi}(\eta)} \int_{{o'}_{l_{t,e}}(\eta)}  \;  {\rm{d}} l_{t,e} {\rm{d}} \phi. 
$ 
Given the domains in Lemma \ref{lm_UAV_2} and dividing \eqref{pro2_eq} by $\pi R_{1}^2$, \eqref{eq_th2_1} is achieved. Thus, we conclude the proof. 
\end{IEEEproof}

\begin{prop} \label{th_UAV_3}
{\rm{Let ${r}\leq R_{1}$, assuming that the mixed LoS/NLoS states exist for link $(s,e)$, and the (quasi-) LoS probability is $\mathbb{P}_{\text{L}}^{s,e}=c_3 l_{t,e}/H +c_4$, the CDF of $\Gamma_2$ conditioning on the UED $e$ being active within the disk $O(t,r)$ is given by
\begin{equation} \label{eq_th3_1}
\begin{split}
F_{\Gamma_5}(y, \eta_{\text{L}}, \eta_{\text{N}},{r}) =& \frac{2}{\pi R_{1}^2}\bigg( c_4 S_1(y,\eta_{\text{L}}, {r})+ \frac{c_3}{3H} S_3(y,\eta_{\text{L}}, {r})+\AND (1-c_4) S_1(y,\eta_{\text{N}}, {r}) -\frac{c_3}{3H} S_3(y,\eta_{\text{N}}, {r})\bigg)
\end{split}
\end{equation}
\begin{equation}
S_3(y,\eta, {r}) =
\begin{cases}
M_3+M_4-M_1-M_2, & \text{ \rm{cases 1, 2} }\\
\pi {r}^3 -2 M_2 + M_3, & \text{ \rm{case 3} } \\
M_1 - M_2 + M_4, & \text{ \rm{cases 4, 14} }  \\
\pi {r}^3,  &\text{ \rm{cases 5, 12, 13, 19} } \\
M_1 - M_3 + M_4, & \text{ \rm{cases 6, 7} }  \\
-M_3, & \text{ \rm{case 8} } \\
-M_1 + M_4, & \text{ \rm{cases 9, 10} }\\
-M_2 + \pi {r}^3, & \text{ \rm{case 11} } \\
-M_2, & \text{ \rm{case 15} } \\
\pi {r}^3 -M_5 -M_6, & \text{ \rm{case 16} }  \\
M_5 +M_6, & \text{ \rm{case 17} } \\
\pi {r}^3/2, & \text{ \rm{case 18} } \\
0, & \text{ \rm {otherwise}}
\end{cases}
\end{equation}
and all the cases with index are given in Table \ref{condition_tabel},
$M_{0}= \frac{\sqrt{-{A} B+l_{t,j}^2}}{-3 {A}^3 /2},
$
$
\frac{2M_1}{M_{0}}= 4 B  {A}  \mathsf{F}\left(\sec ^{-1}\left(\frac{l_{t,j} {r}}{{a_1}/2}\right)|\frac{l_{t,j}^2}{l_{t,j}^2-{A} B}\right)+  \left(8 l_{t,j}^2 -7 B {A} \right)  \mathsf{E}\left(\sec ^{-1}\left(\frac{l_{t,j} {r}}{{a_1}/2}\right)|\frac{l_{t,j}^2}{l_{t,j}^2-{A} B}\right) ,
$
$
M_2=  M_{0}4 B {A}  \mathsf{K}\left(\frac{l_{t,j}^2}{l_{t,j}^2-{A} B}\right)+  M_{0}\left(8 l_{t,j}^2 -7 B {A} \right) \mathsf{E}\left(\frac{l_{t,j}^2}{l_{t,j}^2-{A} B}\right),
$
$
M_3 = M_{0}\left(8 l_{t,j}^2 -7 B {A} \right) \mathsf{E}\left(\sec ^{-1}\left(\frac{-l_{t,j}}{\sqrt{B{A}}}\right)|\frac{l_{t,j}^2}{l_{t,j}^2-{A} B}\right) + M_{0}4 B {A}  \mathsf{F}\left(\sec ^{-1}\left(\frac{-l_{t,j}}{\sqrt{B{A}}}\right)|\frac{l_{t,j}^2}{l_{t,j}^2-{A} B}\right),
$
$
M_4 = {r}^3 \cos ^{-1}\frac{{A} {r}^2+B}{2 l_{t,j} {r}} +\frac{M_{0}}{4r} \left({A}  (2 {a_1}-9 B)+8 l_{t,j}^2 \right) \sqrt{\frac{4 l_{t,j}^2 {r}^2-{a_1}^2}{l_{t,j}^2-{A} B}},
$
$
M_5 = {r}^3 \cos ^{-1}\frac{\left| H^2-l_{t,j}^2\right| }{-2 l_{t,j} {r}},
$
$
M_6 = \frac{ {r} | H^2-l_{t,j}^2| }{8 l_{t,j}^2} \sqrt{4 l_{t,j}^2 {r}^2-\left(H^2-l_{t,j}^2\right)^2}+  \frac{| H^2-l_{t,j}^2| ^3}{16 l_{t,j}^3} \ln \frac{\sqrt{4 l_{t,j}^2 {r}^2-\left(H^2-l_{t,j}^2\right)^2}+2 l_{t,j} {r}} {\left | H^2-l_{t,j}^2 \right|}
$, $\mathsf{K}(\cdot)$ and $\mathsf{F}(\cdot \mid \cdot)$ denote complete and incomplete elliptic integral of the 1st kind \cite{NIST:DLMF}. 
}}
\end{prop}
\begin{IEEEproof}
\begin{equation}\label{pro3_eq}
\begin{split}
\int_0^{2\pi}& \int_0^{{r}}  ({g}(y, \eta_{\text{L}},l_{t,e},\phi)\mathbb{P}_{\text{L}}^{s,e} + {g}(y, \eta_{\text{N}},l_{t,e},\phi)\mathbb{P}_{\text{N}}^{s,e} ){l_{t,e} } \;  {\rm{d}} l_{t,e} {\rm{d}} \phi\\
=&\int_{o_{\phi}(\eta_\text{L})} \int_{o_{l_{t,e}}(\eta_\text{L})}\left( c_3 {l_{t,e} }^2/H+ c_4{l_{t,e} }\right)   {\rm{d}} l_{t,e} {\rm{d}} \phi +\AND \int_{o_{\phi}(\eta_\text{N})} \int_{o_{l_{t,e}}(\eta_\text{N})}\left( -c_3 {l_{t,e} }^2/H+ (1-c_4){l_{t,e} } \right)  {\rm{d}} l_{t,e} {\rm{d}} \phi \\
 = &\frac{c_3 }{ H }\int_{o_{\phi}(\eta_\text{L})} \int_{o_{l_{t,e}}(\eta_\text{L})} l_{t,e}^2 \;  {\rm{d}} l_{t,e} {\rm{d}} \phi + 2 {c_4 S_1(y,\eta_{\text{L}}, {r})} - \AND \frac{c_3  }{ H }\int_{o_{\phi}(\eta_\text{N})} \int_{o_{l_{t,e}}(\eta_\text{N})} l_{t,e}^2 \;  {\rm{d}} l_{t,e} {\rm{d}} \phi + 2{(1-c_4)S_1(y,\eta_{\text{N}}, {r})} 
 \end{split}
\end{equation}
where 
$ \int_{o_{\phi}(\eta)} \int_{o_{l_{t,e}}(\eta)}   l_{t,e}^2  {\rm{d}} l_{t,e} {\rm{d}} \phi+\int_{{o'}_{\phi}(\eta)} \int_{{o'}_{l_{t,e}}(\eta)}  l_{t,e}^2    {\rm{d}} l_{t,e} {\rm{d}} \phi =  \frac{2}{ 3} \pi {r}^3 . 
$ Given Lemma \ref{lm_UAV_2} and dividing \eqref{pro3_eq} by $\pi R_{1}^2$, \eqref{eq_th3_1} is achieved. Thus, we conclude the proof. 
\end{IEEEproof}

\section{Analysis of the SCP}

\subsection{Trend of the SCP with respect to $\frac{P_s}{P_j}$}
\begin{figure}[t]
\centering
\includegraphics[width = 0.481 \textwidth]{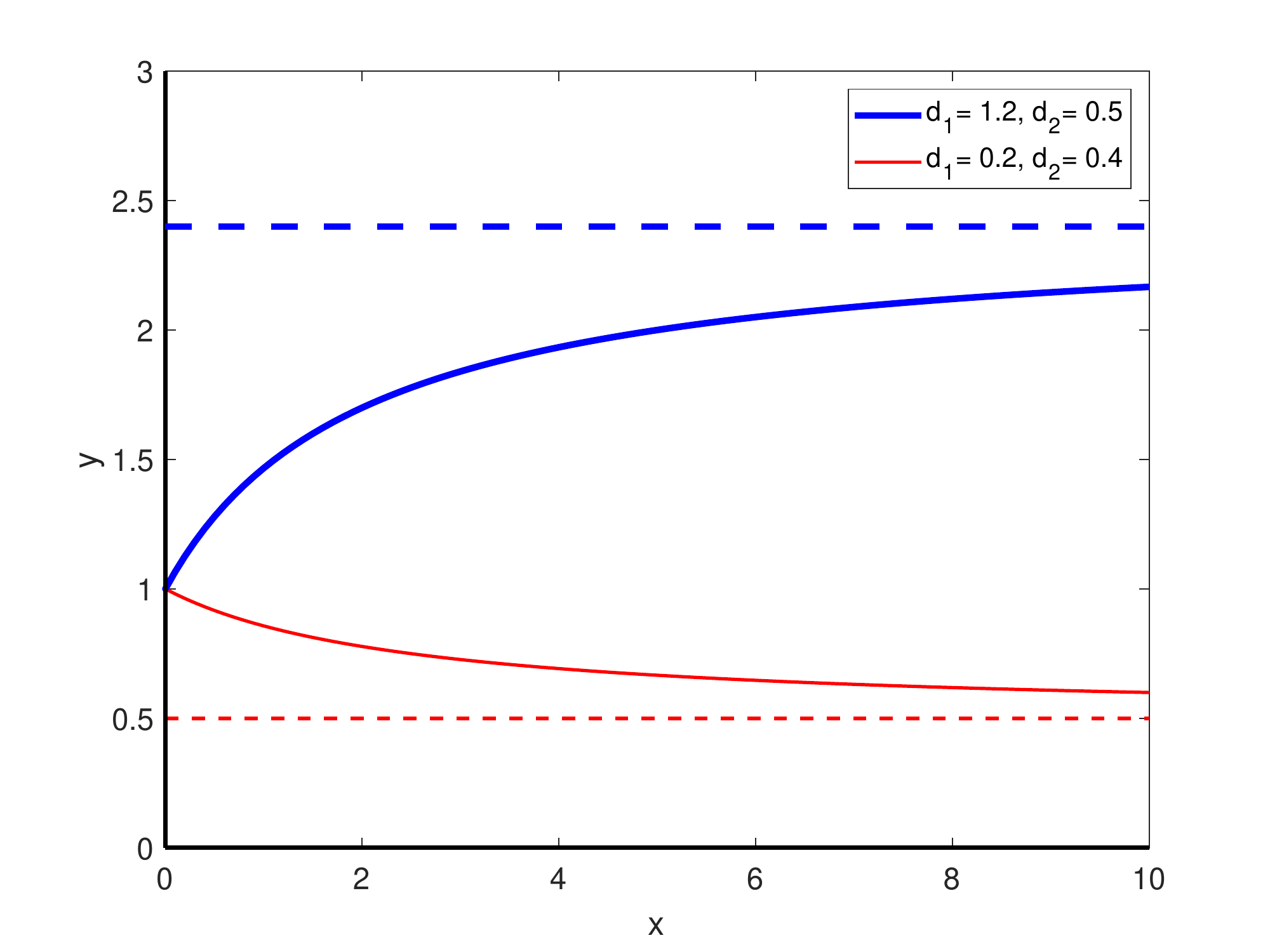}
\caption{The plot of hyperbola \eqref{eq_hyperbolas}. The solid lines are the hyperbolas in the first quadrant and the dashed lines are horizontal asymptotes.}
 \label{fig:hyperbolas}
\end{figure}

This subsection investigates the trend of the SCP with respect to transmitting-to-jamming power ratio $\frac{P_s}{P_j}$ for $d$ and UEDs working in an interference-limited regime. 
Let variable $x = \frac{P_{s}}{P_{j}}$, and random variables ${d_1}= \frac{ l_{{s},{d}}^{-\beta}|h_{{s,d}}|^2}{l_{j,d}^{-\alpha}\eta_{j,d}}$, ${d_2}= \max_{{e} \in \Phi}\left( { \frac{l_{s,e}^{-\alpha} \eta_{s,e}}{l_{j,e}^{-\alpha}}}\right)$, we have   
\begin{equation}\label{eq_hyperbolas}
\begin{split}
y=&\frac{1+\frac{P_{s,d}}{P_{j,d}}}{ 1+\max_{{e} \in \Phi}\left(\frac{P_{s,e}}{P_{j,e}} \right)}  =\frac{1 + {d_1}x  }{1 + {d_2}x}\\
\end{split}
\end{equation}
for $x\geq 0$. As shown in Fig. \ref{fig:hyperbolas}, $y$ is a hyperbola for $d_1, d_2>0$. Its horizontal asymptote is $y = d_1/d_2$. A special point that the curve passes through the x-axis is $(-1/d_1,0)$, and the vertical asymptote of the hyperbola is given by $x = -1/d_2$. Clearly, when $x=0$, $y=1$ and the achievable secrecy rate is zero. 

For $ {d_1}> {d_2}$, the hyperbola $y$ is monotonically increasing in $x$, and if $ {d_1} < {d_2}$, $y$ is monotonically decreasing on $x$. Note that \eqref{security_con} defines the statistical comparison between \eqref{eq_hyperbolas} and $2^{\mathcal{R}_t}$ where $\mathcal{R}_t\geq 0$. For ${d_1} < {d_2}$, $y$ is always less than one. Hence, it is smaller than $2^{\mathcal{R}_t}$. Therefore, when ${d_1} < {d_2}$, neither increasing nor decreasing $P_s/P_j$ will affect the results of the comparison between \eqref{eq_hyperbolas} and $2^{\mathcal{R}_t}$. Consequently, the monotonicity of \eqref{security_con} is solely decided by ${d_1}> {d_2}$. As a result, the SCP is monotonically increasing as $P_s/P_j$ increases. 

Meanwhile, when $\mathcal{R}_t=0$, neither increasing nor decreasing $P_s/P_j$ will affect the results of the comparison between \eqref{security_con} and $2^{\mathcal{R}_t}$ because \eqref{eq_hyperbolas} is always larger than one for ${d_1} > {d_2}$, while \eqref{eq_hyperbolas} is always smaller than one for ${d_1} < {d_2}$. 
To conclude, the analysis above shows that for a given $P_s$, the less $P_j$ the better the SCP in the interference-limited environment. As we will see in the following section, deploying the UAV jammer to a carefully chosen position would benefit the communication networks than the case without the UAV jammer. 

\subsection{Trend of the SCP with respect to jammer's position}
This subsection investigates the trend of the SCP with respect to jammer's position when all ATG links are in a pure LoS state and $H^2 >> (R_{1}+ l_{s,d})^2$. 
An important aspect of UAV networks is the high
likelihood of having significantly stronger LoS components than the reflected multipath components in some deployment scenarios (especially in suburban and rural morphologies) \cite{chetlur2017downlink}.
Hence, it is very important to study the performance of the SCP for the ATG links in a pure LoS state. According to \eqref{eq_fitting_32}, for ATG links in a pure LoS state, as $H$ increases, they will remain in a pure LoS state.

Given $l_{j,d}^2 = l_{t,j}^2 + l_{s,d}^2 - 2l_{t,j} l_{s,d} \cos\left( \varphi_{j'} - \varphi_{d}\right) + H^2$ and $l_{s,e}^2 =H^2+l_{t,e}^2$, when $H^2 >> (R_{1}+ l_{s,d})^2$ and $R_{1}\geq l_{t,j}$, we have $l_{j,d}\approx l_{s,e} \approx H $, and 
\begin{equation} \label{eq_s1_11}
\begin{split}
&\frac{1+\frac{P_{s,d}}{P_{j,d}+N_0}}{ 1+\max_{{e} \in \Phi}\left(\frac{P_{s,e}}{P_{j,e}+N_e} \right)}\approx \frac{1+ \frac{P_{s} l_{{s},{d}}^{-\beta}|{h}_{{s,d}}|^2}{P_{j}H^{-{\alpha}}\eta_{L} +N_0}}{ 1+ \frac{ P_{s}\eta_{L}}{H^{\alpha}} \max_{{e} \in \Phi} \left( \frac{1}{P_{j}l_{j,e}^{-\alpha}+N_e} \right) }.\\
\end{split}
\end{equation}

According to \eqref{eq_s1_11}, the performance of the SCP is related to $\max_{{e} \in \Phi} {l_{j,e}}$. For a UED $e$ uniformly distributed within a disk, the PDF of $l_{e,j}$ conditoning on $l_{t,j}$ is proportional to the circular arc of the circle $O(j, l_{j,e})$ that is enclosed by the disk $O(t,R_{1})$ \cite{tang2017distribution}. Assuming that for a given $l_{t,j}$, the domain of $l_{j,e}$ with non-zero density is $[0, Z_1]$, where $Z_1>0$. Then, as $l_{t,j}$ increases by $\delta>0$ and $j$ still stays within the circle $O(t,R_{1})$. The new non-zero domain of $l_{j,e}$ is $[0, Z_1 + \delta]$, and the density of $l_{j,e}$ is decreasing in $l_{t,j}$ for any $l_{j,e}\in [0, Z_1]$. Thus, the mean of $l_{j,e}$ increases as $l_{t,j}$ increases. As a result, when the number of the UEDs is one, placing UAV jammer $j$ to $t$ (i.e., $l_{t,j}=0$) will provide the optimum SCP for $H^2 >> (R_{1}+ l_{s,d})^2$. 

Meanwhile, for multiple UEDs, we have the following discussion.
Let $n$ i.i.d. random variables $X_1,...,X_n$ denote the distances from $n$ UEDs to $j$, the CDF of $Y=\max(X_1,...,X_n)$ conditioning on $l_{t,j}$ is given by
\begin{align}
P(Y\leq y \mid l_{t,j})  
            &\stackrel{i.i.d}{=}  P(X_1\leq y \mid l_{t,j})^n
\end{align}
and the corresponding PDF is written as
\begin{align}\label{eq_sp_3}
f_Y(y\mid l_{t,j}) = n(P(X_1\leq y \mid l_{t,j}))^{n-1}f_{X_1}(y\mid l_{t,j}) 
\end{align}
where $f_{X_1}(y\mid l_{t,j})$ is the conditional PDF of $X_1$. For an infinitesimal $ \Delta \delta$, when the upper bound of the domain of $l_{t,j}$ with non-zero density increases from $Z_1$ to $Z_1 +  \Delta \delta$ by moving $j$ away from $t$, we have 
\begin{align}\label{eq_sp_1}
f_Y(Z_1 + \Delta \delta \mid l_{t,j} + \Delta \delta) = nf_{X_1}(Z_1 +  \Delta \delta \mid l_{t,j}+ \Delta \delta )>0
\end{align}
since $P(X_1 \leq Z_1+  \Delta \delta \mid l_{t,j}+  \Delta \delta) =1$.
Meanwhile, as a result of the previous discussion for one UED, we have both $f_{X_1}(y\mid l_{t,j}) \geq f_{X_1}(y\mid l_{t,j}+\Delta \delta)$ and $P(X_1\leq y \mid l_{t,j})\geq P(X_1\leq y \mid l_{t,j}+ \Delta \delta)$ for $y\in [0, Z_1]$. According to \eqref{eq_sp_3}, 
\begin{align}\label{eq_sp_2}
f_Y(y\mid l_{t,j}+ \Delta \delta) \leq f_Y(y\mid l_{t,j}) 
\end{align}
for $y\in [0, Z_1]$. As a result of \eqref{eq_sp_1} and \eqref{eq_sp_2}, the mean of $\max_{e \in \Phi} l_{j,e}$ increases as $l_{t,j}$ increases. Thus, placing UAV jammer $j$ at $t$ will obtain the optimum SCP for $H^2 >> (R_{1}+ l_{s,d})^2$.

\section{Simulation Results}

To validate the derivation and the analysis of the SCP with respect to different variables, simulations based on LTE parameters were developed. The parameters used in the simulation are given in Table \ref{params_label} unless otherwise specified. The receiver $d$ and UEDs are assumed to have the same noise power.  The simulation results are obtained by averaging over $1 \times 10^5$ independent Monte Carlo trials.

\begin{table}[t]
\renewcommand{\arraystretch}{0.6}
\centering
\caption{System parameters}
\label{params_label}
\begin{tabular}{ll}
\hline
Parameter                            & Value                 \\ \hline
Environment 						& Suburban\\
Disk radius $R_{1}$ 				&$500$ m\\
Receiver location $d$ 				&$(100,0,0)$\\
Coding gain							& 0 dB
\\
Tx/Rx antenna gain                 & 0 dBi                   \\
Receiver noise figure              & 9 dB                  
\\
Carrier frequency                  & 2.0 GHz                  \\
Spectrum allocation               	& 20 MHz                  \\
Duplex mode					        & Half duplex                  \\
Thermal noise power density	        & -174 dBm/Hz                   \\
Path loss exponent $\beta$         & 3 \\
Transmit power $P_s$   				& $0.1$ W                 \\
Target secrecy rate ${\mathcal{R}_t}$                         & 1 bit/s/Hz \\
Number of UEDs $n$        		    & 1\\ \hline
\end{tabular}
\end{table}

\begin{figure}[t]
\centering
\includegraphics[width = 0.481 \textwidth]{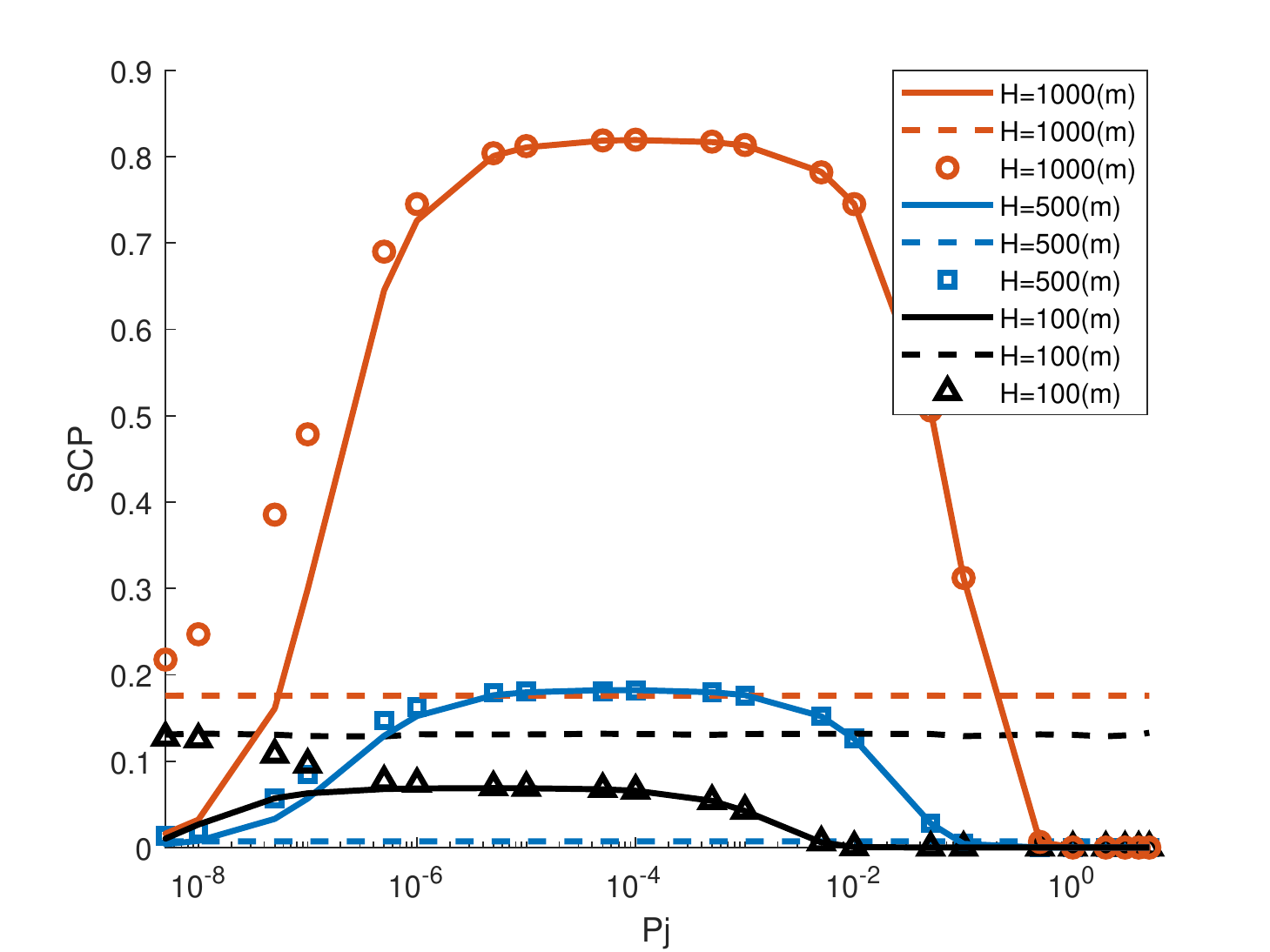}
\caption{SCP \textit{vs.} ground location of UAV jammer for different $P_j$ in watt as given by the figure legends. For the same color, the solid line represents the numerical results with the UAV jammer, the markers represent the simulation results with the UAV jammer and the dashed line represents the simulation results without the UAV jammer. $j=(0,0, 500)$.}
 \label{fig:UAV_5}
\end{figure}

\begin{figure}[t]
\centering
\includegraphics[width = 0.481 \textwidth]{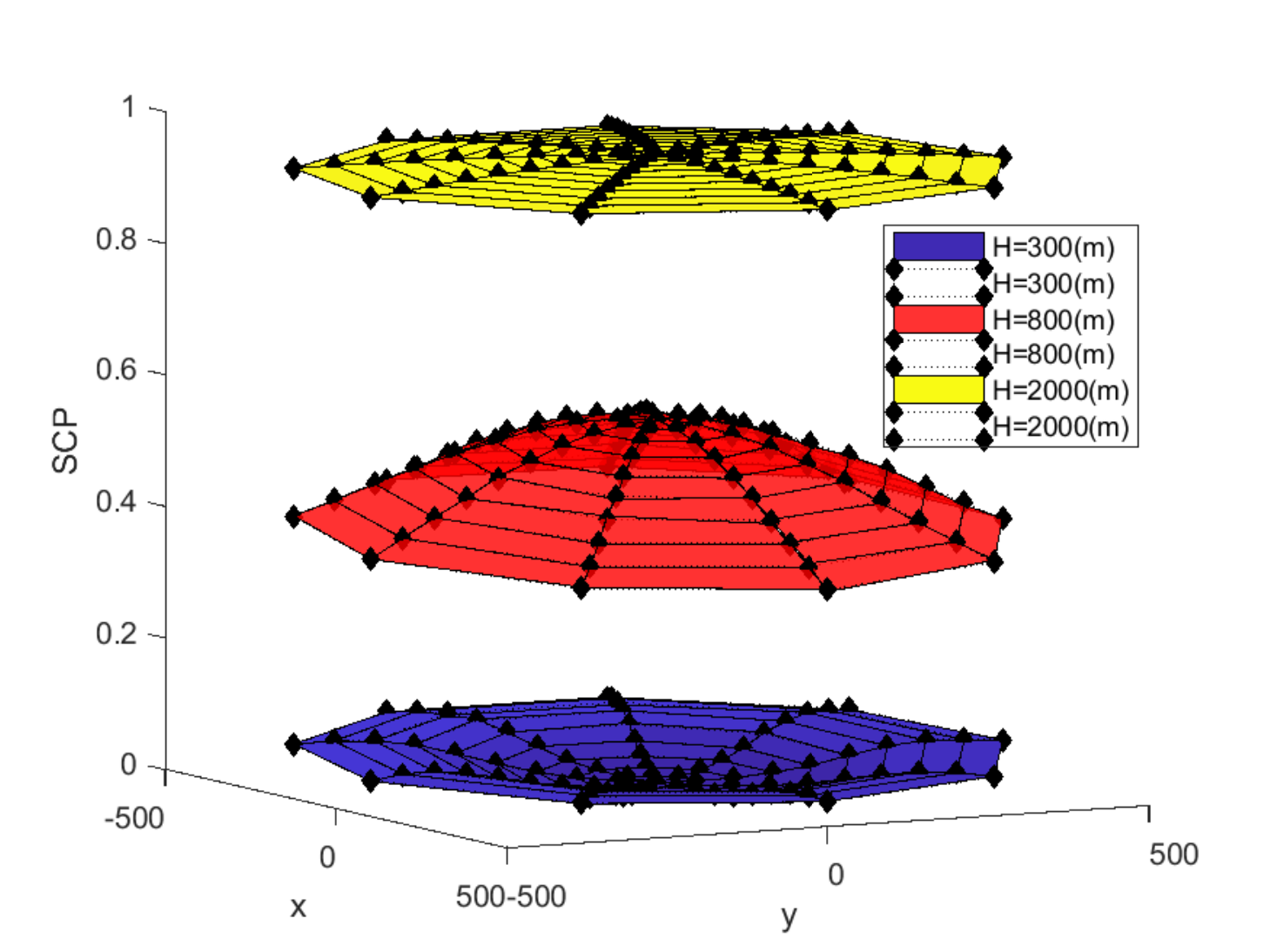}
\caption{SCP \textit{vs.} location of UAV jammer, where $P_j = 0.01 $ W. The mesh grids represent simulation results, and the markers represent numerical results.}
 \label{fig:UAV_4}
\end{figure}

\begin{figure}[t]
\centering
\includegraphics[width = 0.481 \textwidth]{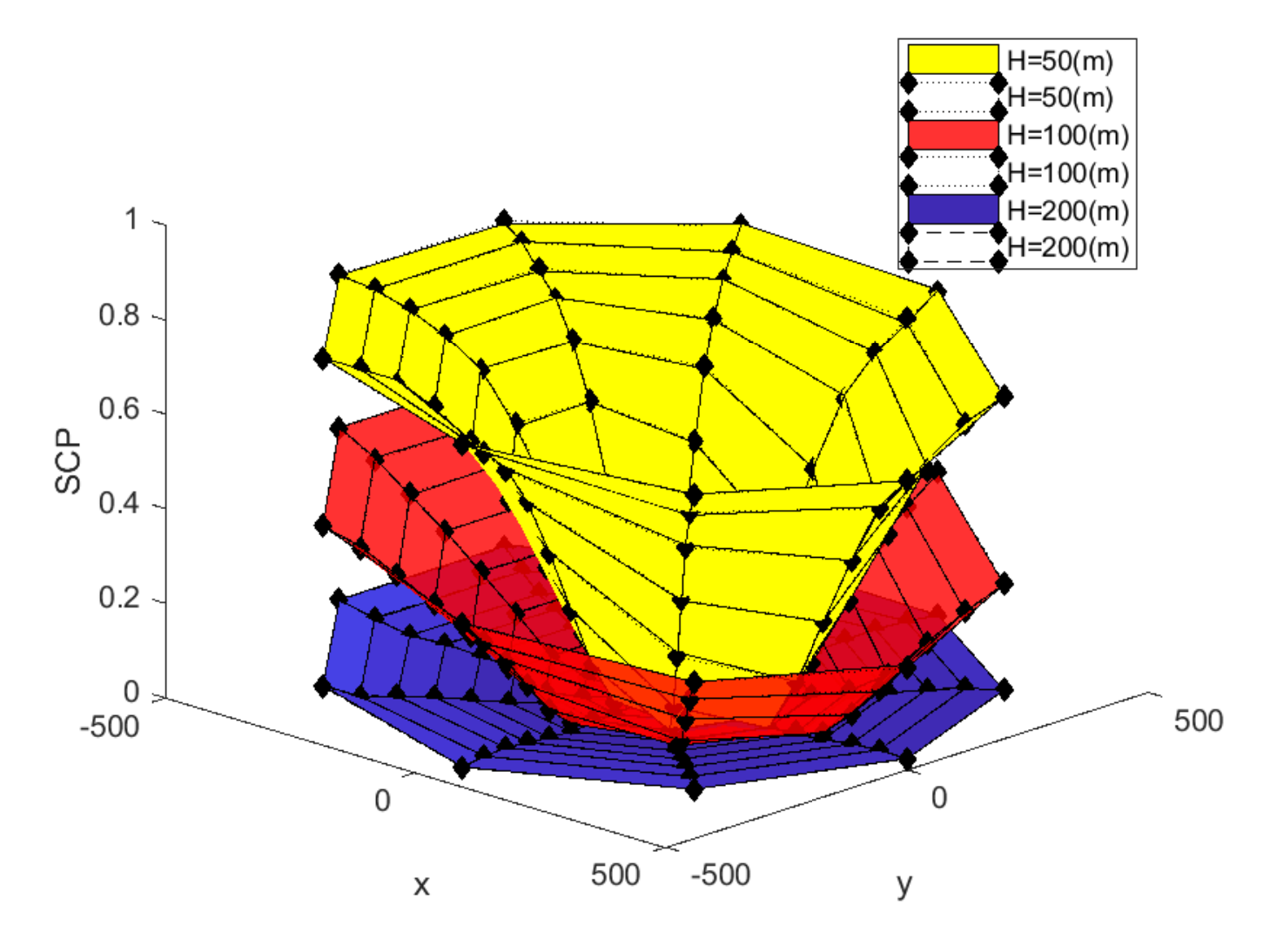}
\caption{SCP \textit{vs.} location of UAV jammer, where $P_j = 0.01 $ W. The mesh grids represent simulation results, and the markers represent numerical results.}
 \label{fig:UAV_7}
\end{figure}

Fig. \ref{fig:UAV_5} shows the performance of the SCP in the simulation for UEDs at different heights. The numerical results in this section are obtained by evaluating \eqref{SSP_lb} for the same parameters as the simulations. One can easily see that the numerical results act as the lower bound to the simulation results of the SCP when $P_j$ is small. 
The simulation curves match well with the numerical results as $P_j$ reaches the interference-limited regime. In this regime, the performance of the SCP is decreasing as $P_j$ increases. This can be easily explained by the discussion on $P_s/P_j$ in \eqref{eq_hyperbolas}. Meanwhile, the same figure also shows that introducing a UAV jammer to the ground communications could change the performance of the SCP significantly. %
On the one hand, for $H=500$ m and $1000$ m, the SCP increases as $P_j$ increases until around $10^{-3}$ W and then starts to decrease until the SCP reaches zero. Compared to the similar settings without the UAV jammer, introducing a UAV jammer with proper $P_j$  would allow the ground communications to achieve a higher SCP. For $H=500$ m and $1000$ m, the ground communications with a UAV jammer allow the maximum of 20\% and 60\% increase in SCP, respectively, over the case without a UAV jammer as given by the dashed lines. On the other hand, for $H=100$ m, the SCP increases as $P_j$ increases. By having a UAV jammer in this height, the SCP of the ground communications will be affected severely. 
Meanwhile, notice the case without the UAV jammer, the achieved SCP in dashed lines show that the SCP at $H=100$ m is higher than the one at $H=500$ m, while this trend is reversed when comparing $H=500$ m to $H=1000$ m. It is because increasing $H$ will cause not only a higher LoS probability from the transmitter to a UED but also a longer receiving distance (i.e., a higher path loss). In general, the ground communications prefer to have NLoS links towards the UEDs and have a maximum distance from it, however, these two objectives conflicts with each other. If the LoS probability from the transmitter to a UED contributes more than the path loss does to the SCP as $H$ increases, the SCP will decrease. Otherwise, the SCP will increase.

Fig. \ref{fig:UAV_4} depicts the simulation results on the SCP when all ATG links are in a pure LoS state.  When $H=2000$, the performance of the SCP is optimum when placing the UAV jammer at $(0,0,H)$. This can be explained by the discussion on \eqref{eq_s1_11}. The performance curves from $H=300$ to $H=2000$ show that, for the UAV jammer with the same x and y coordinates, the SCP increases as $H$ increases. This can also be explained by \eqref{eq_s1_11}, where the right term decreases as $H$ increases for the ATG links in a pure LoS state. 
Meanwhile, Fig. \ref{fig:UAV_7} gives the cases without the requirement on a pure LoS state for ATG links. In general, the trend of the SCP with respect to the height of UAVs is hard to capture, because increasing $H$ will not only cause a higher LoS probability but also decreases the jamming and received signal powers. A higher LoS probability from $s$ to UED will cause the SCP to drop while a smaller jamming power will cause the SCP to increase. Meanwhile, the convex surface shows that the jammer should move away from the receiver $d$ in order to achieve a higher SCP.
Furthermore, by observing both Fig. \ref{fig:UAV_4} and Fig. \ref{fig:UAV_7}, the numerical results of the SCP matching with the simulation results suggest that both receiver $d$ and UEDs are working in the interference-limited regime for $P_j=0.01$ W. By increasing $H$ from $50$ m to $200$m, the SCP at the given x and y coordinates decreases first and then increases.

\section{Conclusion}
In this paper, we investigated the secrecy performance of the ground link in the presence of randomly deployed non-colluding UEDs. A piecewise fitting model has been proposed to characterize the LoS probability of an ATG link with higher accuracy than the existing sigmoid fitting. The SCP has been formulated based on the piecewise fitting model so that the formulation become tractable. 
The performance of the SCP with respect to transmitting-to-jamming power ratio, the height of the UAVs and the UAV jammer location has been investigated. The effective jamming power and location of the UAV jammer have been pointed out. The model and analysis frameworks presented in this paper will help to facilitate the analysis of the UAV-aided communications.

\section*{Appendix}
\begin{lemma}\label{lm_UAV_2}
{\rm For a UED $e$ which is uniformly distributed within the disk $O(t,{r})$, denote $\phi = \varphi_{j'}- \varphi_{e'}$ and $\eta$ as the excess path loss, the domains of $\phi$ and $l_{t,e}$ so that $1+\frac{(H^2+l_{t,e}^2)^\frac{-\alpha}{2}\eta P_s}{\left(l_{t,j}^2+l_{t,e}^2-2 l_{t,j} l_{t,e}\cos \phi\right)^\frac{-\alpha}{2}P_j} \leq y$ is denoted as  $\{{o}_{\phi}(\eta ), {o}_{l_{t,e}}(\eta)\}$, while the domains of $\phi$ and $l_{t,e}$ so that $1+\frac{(H^2+l_{t,e}^2)^\frac{-\alpha}{2}\eta P_s}{\left(l_{t,j}^2+l_{t,e}^2-2 l_{t,j} l_{t,e}\cos \phi\right)^\frac{-\alpha}{2}P_j} > y$ is denoted as $\{{o'}_{\phi}(\eta ), {o'}_{l_{t,e}}(\eta) \}$. Let $\mathcal{F}_1 = {\cos^{-1}} \frac{-\sqrt{B{A}}}{l_{t,j}}$, $\mathcal{F}_2 = {\cos^{-1}} \frac{{r}^2{A} + B}{2 {r} l_{t,j}}$, ${A} = 1-\left((y-1)\frac{P_j}{\eta P_s}\right)^\frac{2}{\alpha}$, $B= l_{t,j}^2- (1-{A}) H^2$. $x_1 = \frac{ l_{t,j} \cos \phi + \sqrt{\left( l_{t,j} \cos \phi \right )^2 -  B{A}}}{{A}}$, and $x_2 = \frac{ l_{t,j} \cos \phi - \sqrt{\left( l_{t,j} \cos \phi \right )^2 -  B{A}}}{{A}}$.
Based on different ${A}$ and $B$, these domains are given as follows.

For ${A}<0, B \leq 0$, we have ${o'}_{l_{t,e}}(\eta) = [x_1, x_2]$ and
\begin{equation}\label{eq_codition_A111_2}
{o'}_{\phi}(\eta )=
\begin{cases}
\left[\mathcal{F}_1, \mathcal{F}_2\right], &\text{if } {r}^2{A}\leq B,  \frac{-{r}^2{A} - B}{2 {r}} \leq l_{t,j}\\
\left[\mathcal{F}_1, \pi\right], &\text{if } {r}^2{A} < B, \sqrt{B{A}} \leq l_{t,j} < \frac{-{r}^2{A} - B}{2 {r}} \\
 \{\emptyset\},  &\text{otherwise;}
\end{cases}
\end{equation}

or $ {o'}_{l_{t,e}}(\eta) = [x_1, {r}]$ and

\begin{equation}\label{eq_codition_A111_21}
{o'}_{\phi}(\eta )=
\begin{cases}
\left(\mathcal{F}_2, \pi\right], & \text{if } {r}^2{A}  \leq B,  \frac{-{r}^2{A} - B}{2 {r}} \leq l_{t,j} \\
\left[\mathcal{F}_2, \pi \right], &\text{if } {r}^2{A} > B,  \frac{-{r}^2{A} - B}{2 {r}} \leq l_{t,j} \\
\{\emptyset\}, & \text{otherwise. } \\
\end{cases}
\end{equation}

For ${A}>0, B \geq 0$, we have ${o}_{l_{t,e}}(\eta) = [x_2, x_1]$ and
\begin{equation}\label{eq_lemma_1_3}
{o}_{\phi}(\eta )=
\begin{cases}
\left[\mathcal{F}_2, \pi- \mathcal{F}_1\right], 
 &\text{if } {r}^2{A}\geq B, l_{t,j} \geq \frac{{r}^2{A} + B}{2 {r}}\\
\left[0, \pi- \mathcal{F}_1\right],  &\text{if }  {r}^2{A} > B, \sqrt{B{A}} \leq l_{t,j} < \frac{{r}^2{A} + B}{2 {r}} \\
\{\emptyset\},  &\text{otherwise;}
\end{cases}
\end{equation}

or ${o}_{l_{t,e}}(\eta) = [x_2, {r}]$ and
\begin{equation}\label{eq_lemma_1_4}
{o}_{\phi}(\eta )=
\begin{cases}
\left[0,  
\mathcal{F}_2 \right],  &\text{if } {r}^2{A}  \leq B,  \frac{{r}^2{A} + B}{2 {r}} \leq l_{t,j}  \\
\left[0, \mathcal{F}_2\right),  &\text{if } {r}^2{A} > B, \frac{{r}^2{A} + B}{2 {r}} \leq l_{t,j}  \\
\{\emptyset\}, & \text{otherwise.}
\end{cases}
\end{equation}

For ${A}<0, B > 0$, we have ${o'}_{l_{t,e}}(\eta) = [0, {r}]$ and
\begin{equation}\label{eq_lemma_1_5}
{o'}_{\phi}(\eta )=
\begin{cases}
[0,\pi],  &\text{if } -{r}^2{A} < B, l_{t,j}< \frac{{r}^2{A} + B}{2 {r}} \\
\left(\mathcal{F}_2, \pi \right], & \text{if }  l_{t,j}\geq  \frac{\left |{r}^2{A} + B\right|}{2 {r}}  \\
\{\emptyset\}, & \text{otherwise;}
\end{cases}
\end{equation}

or ${o'}_{l_{t,e}}(\eta) =[0,x_2]$ and
\begin{equation}\label{eq_lemma_1_6}
{o'}_{\phi}(\eta )=
\begin{cases}
[0,\pi],  &\text{if } -{r}^2{A}> B, l_{t,j}< \frac{-{r}^2{A} - B}{2 {r}} \\
\left[0, \mathcal{F}_2\right], & \text{if } l_{t,j}\geq \frac{\left| {r}^2{A} + B\right|}{2 {r}} \\
\{\emptyset\}, & \text{otherwise.}
\end{cases}
\end{equation}

For ${A}>0, B < 0$, we have
 ${o}_{l_{t,e}}(\eta) =[0, {r}]$ and
\begin{equation}\label{eq_lemma_1_7}
{o}_{\phi}(\eta )=
\begin{cases}
[0,\pi],  &\text{if } -{r}^2{A} > B,  l_{t,j}< \frac{-{r}^2{A} - B}{2 {r}} \\
\left[ 0, \mathcal{F}_2 \right], & \text{if } l_{t,j}\geq \frac{|{r}^2{A} + B |}{2 {r}} \\
\{\emptyset\}, & \text{otherwise;}
\end{cases}
\end{equation}

or ${o}_{l_{t,e}}(\eta) =[0, x_1]$ and
\begin{equation}\label{eq_lemma_1_8}
{o}_{\phi}(\eta )=
\begin{cases}
[0,\pi],  &\text{if } -{r}^2{A} < B, l_{t,j}< \frac{{r}^2{A} + B}{2 {r}} \\
\left[\mathcal{F}_2, \pi\right], & \text{if } l_{t,j}\geq \frac{|{r}^2{A} + B|}{2 {r}} \\
\{\emptyset\}, & \text{otherwise.}
\end{cases}
\end{equation}

For ${A}=0$, we have
${o}_{l_{t,e}}(\eta) = \left[ \frac{\left | l_{t,j}^2 -H^2\right|}{2 l_{t,j}  } ,  {r}\right]$ and

\begin{equation}\label{eq_new_region_A1}
{o}_{\phi}(\eta )=
\begin{cases}
\left[0, {\cos^{-1}} \left(\frac{ l_{t,j}^2 -H^2}{2 l_{t,j}  l_{t,e}}\right)\right], & \text{if }   \frac{\left | l_{t,j}^2 -H^2\right| }{2 l_{t,j}  }  \leq {r}\\
\{\emptyset\}, & \text{otherwise;}
\end{cases}
\end{equation}

or ${o}_{l_{t,e}}(\eta) = \left[0,  \frac{ -l_{t,j}^2 +H^2}{2 l_{t,j}  } \right)$ and
\begin{equation}\label{eq_new_region_A2}
{o}_{\phi}(\eta )=
\begin{cases}
[0,\pi],  &\text{if } {r}\geq  \frac{ -l_{t,j}^2 +H^2}{2 l_{t,j}  }  \\
\{\emptyset\}, & \text{otherwise;}
\end{cases}
\end{equation}

or ${o}_{l_{t,e}}(\eta) = \left[0, {r}\right]$ and

\begin{equation}\label{eq_new_region_A3}
{o}_{\phi}(\eta )=
\begin{cases}
[0,\pi],  &\text{if } {r}<  \frac{ -l_{t,j}^2 +H^2}{2 l_{t,j}  }  \\
\{\emptyset\}, & \text{otherwise.}
\end{cases}
\end{equation}
}

\end{lemma}

\begin{IEEEproof}
For ${A}\neq 0$, the corresponding solutions of $l_{t,e}$ in $1+\frac{(H^2+l_{t,e}^2)^\frac{-\alpha}{2}\eta P_s}{\left(l_{t,j}^2+l_{t,e}^2-2 l_{t,j} l_{t,e}\cos \phi\right)^\frac{-\alpha}{2}P_j} = y$ are denoted as
\begin{equation}\label{eq_x_1_x_2}
\begin{split}
x_1 = \frac{ l_{t,j} \cos \phi + \sqrt{\left( l_{t,j} \cos \phi \right )^2 -  B{A}}}{{A}}\\
x_2 = \frac{ l_{t,j} \cos \phi - \sqrt{\left( l_{t,j} \cos \phi \right )^2 -  B{A}}}{{A}}
\end{split}
\end{equation}
where $x_1 \leq x_2$ for ${A}<0$, while $x_2 \leq x_1$ for ${A}>0$. 
For $x_1$ and $x_2$ to exist, 
\begin{equation}\label{eq_codition_A1_con}
{\left( l_{t,j} \cos \phi \right )^2 -  B{A}}\geq 0.
\end{equation}
As a result, $\cos \phi \geq \frac{\sqrt{B{A}}}{l_{t,j}}$ or $\cos \phi \leq -\frac{\sqrt{B{A}}}{l_{t,j}}$ for $B{A}\geq 0$, and ${\left( l_{t,j} \cos \phi \right )^2 }\geq 0 $ for $B{A}< 0$. Since $\phi = \varphi_{j'}- \varphi_{e'}$ and the impact from $j$ to $e$ for $\phi \in [0,\pi]$ is the same as the case for $\phi \in [\pi,2\pi]$. Thus, we only consider $\phi \in [0,\pi]$ in the following analysis. The domains of $\phi$ in \eqref{eq_codition_A1_con} are given by
\begin{equation}\label{eq_codition_A1_1}
\phi \leq \pi - \mathcal{F}_1, \text{ for } l_{t,j}\geq \sqrt{B{A}}, B{A}\geq 0,\\
\end{equation}
\begin{equation}\label{eq_codition_A1_2}
\phi \geq \mathcal{F}_1, \text 
{ for } l_{t,j}\geq \sqrt{B{A}} , B{A}\geq 0, \\
\end{equation}
\begin{equation}\label{eq_1}
\phi \in [0, \pi], \text{ for } B{A}< 0. 
\end{equation}

Meanwhile, $1+\frac{(H^2+l_{t,e}^2)^\frac{-\alpha}{2}\eta P_s}{\left(l_{t,j}^2+l_{t,e}^2-2 l_{t,j} l_{t,e}\cos \phi\right)^\frac{-\alpha}{2}P_j} \leq y$ gives
\begin{equation} 
\begin{split}
A l_{t,e}^2 - 2 l_{t,j} \cos (\phi ) l_{t,e} + l_{t,j}^2 - (1-A) H^2 \leq 0.
\end{split}
\end{equation}

Apparently, when $A<0$, $l_{t,e} \geq x_2$ or $l_{t,e} \leq x_1$; when $A>0$, $x_2 \leq l_{t,e} \leq x_1$. To simplify analysis, we investigate $x_1 \leq l_{t,e} \leq x_2$ for $A<0$ and $x_2 \leq l_{t,e} \leq x_1$ for $A>0$. As a result, 
the former is the case for $A<0$ and
\begin{equation} 
\begin{split}
A l_{t,e}^2 - 2 l_{t,j} \cos (\phi ) l_{t,e} + l_{t,j}^2 - (1-A) H^2 \geq 0.
\end{split}
\end{equation}

{\bf {For ${A}<0, B \leq 0$:}}
Assuming that \eqref{eq_codition_A1_con} holds, comparing $x_1$ with $0$ gives
\begin{equation}\label{eq_codition_A1_bd}
\begin{split}
\phi \in \left[\frac{\pi}{2}, \pi \right], \text{ for } x_1 \geq 0; \text{ and }
\phi \in \left[0,\frac{\pi}{2} \right],\text{ for } x_1 <0.
\end{split}
\end{equation}

Assuming that \eqref{eq_codition_A1_con} holds, comparing $x_2$ with $0$ and considering that $x_1 \leq x_2$ for ${A}<0$, we have
\begin{equation}\label{eq_codition_A1_bd_ex}
\begin{split}
\phi \in \left[\frac{\pi}{2}, \pi \right], & \,x_2 \geq 0.
\end{split}
\end{equation}

Comparing $x_1$ with ${r}$, we have
\begin{equation}\label{eq_codition_A2_1}
\begin{split}
x_1\leq {r} \Rightarrow \sqrt{\left( l_{t,j} \cos \phi \right )^2 -  B{A}} \geq {r}{A}- l_{t,j} \cos \phi.\\
\end{split}
\end{equation}

If 
\begin{equation}\label{eq_ex_5}
{r}{A}- l_{t,j} \cos \phi\geq 0,
\end{equation} 
then
\begin{equation}\label{eq_codition_A2}
\phi \in \begin{cases}
[\mathcal{F}_3, \pi], & \text{if } l_{t,j} \geq -{r}{A} \\
\{\emptyset\}, & \text{if }  l_{t,j} < -{r}{A} \\
\end{cases}
\end{equation}
where $\mathcal{F}_3 = {\cos^{-1}} \frac{{r}{A}}{l_{t,j}}$.

Assuming that \eqref{eq_codition_A1_con} and \eqref{eq_ex_5} hold, solving \eqref{eq_codition_A2_1} gives
\begin{equation}\label{eq_codition_A3}
\phi \in \begin{cases}
[\mathcal{F}_2, \pi], & \text{if } l_{t,j}\geq \frac{-{r}^2{A} - B}{2 {r}} \\
\{\emptyset\}, & \text{otherwise. }  \\
\end{cases}
\end{equation}

Combining \eqref{eq_codition_A2} and \eqref{eq_codition_A3} with intersections yields
\begin{equation}\label{eq_codition_A4}
\phi \in \begin{cases}
[\mathcal{F}_3, \pi], & \text{if } {r}^2{A} \leq B, \, l_{t,j} \geq {-{r}{A}} \\
[\mathcal{F}_2, \pi], & \text{if }  {r}^2{A} > B, \,  l_{t,j} \geq \frac{-{r}^2{A} - B}{2 {r}} \\
\{\emptyset\}, & \text{otherwise. } \\
\end{cases}
\end{equation}

Combining \eqref{eq_codition_A4} and \eqref{eq_codition_A1_2} with intersections yields
\begin{equation}\label{eq_codition_A6}
\phi \in \begin{cases}
[\mathcal{F}_3, \pi], & \text{if } {r}^2{A} \leq B, \, l_{t,j} \geq {-{r}{A}} \\
[\mathcal{F}_2, \pi], & \text{if }  {r}^2{A} > B, \, l_{t,j} \geq \frac{-{r}^2{A} - B}{2 {r}} \\
\{\emptyset\}, & \text{otherwise. } \\
\end{cases}
\end{equation}

If ${r}{A}- l_{t,j} \cos \phi < 0$, then
\begin{equation}\label{eq_codition_A1111_22}
\phi \in \begin{cases}
[0, \pi], & \text{if } l_{t,j} < -{r}{A}\\
[0, \mathcal{F}_3), & \text{if } l_{t,j} \geq -{r}{A}. \\
\end{cases}
\end{equation}

Combining \eqref{eq_codition_A1111_22} and \eqref{eq_codition_A1_2} with intersections yields
\begin{equation}\label{eq_codition_A1111_22_1111}
\phi \in \begin{cases}
[\mathcal{F}_1, \pi], & \text{if } {r}^2{A} < B,  \sqrt{B{A}} \leq l_{t,j} <-{r}{A}\\
[\mathcal{F}_1, \mathcal{F}_3), & \text{if } {r}^2{A} < B,  l_{t,j} \geq -{r}{A} \\
\{\emptyset\}, & \text{otherwise.}
\end{cases}
\end{equation}

The unions of \eqref{eq_codition_A6} and \eqref{eq_codition_A1111_22_1111} give

\begin{equation}\label{eq_codition_A1111_22_1111_1}
\phi \in \begin{cases}
[\mathcal{F}_2, \pi], & \text{if }  {r}^2{A} > B, \, l_{t,j} \geq \frac{-{r}^2{A} - B}{2 {r}} \\
[\mathcal{F}_1, \pi], & \text{if } {r}^2{A} \leq B,  \sqrt{B{A}} \leq l_{t,j} \\
\{\emptyset\}, & \text{otherwise.}
\end{cases}
\end{equation}

Comparing ${r}$ with $x_2$, 
\begin{equation}\label{eq_codition_A21}
\begin{split}
x_2> {r} \Rightarrow \sqrt{\left( l_{t,j} \cos \phi \right )^2 -  B{A}} > -{r}{A}+ l_{t,j} \cos \phi.\\
\end{split}
\end{equation}

If 
\begin{equation}\label{eq_ex_4}
-{r}{A}+ l_{t,j} \cos \phi\geq 0,  
\end{equation}
then
\begin{equation}\label{eq_codition_A22}
\phi \in \begin{cases}
[0, \pi], & \text{if } l_{t,j} < -{r}{A}\\
[0, \mathcal{F}_3], & \text{if } l_{t,j} \geq -{r}{A}. \\
\end{cases}
\end{equation}

Assuming that \eqref{eq_codition_A1_con} and \eqref{eq_ex_4} hold, solving \eqref{eq_codition_A21} gives
\begin{equation}\label{eq_codition_A23}
\phi \in \begin{cases}
(\mathcal{F}_2, \pi],  &\text{if } l_{t,j}\geq \frac{-{r}^2{A} - B}{2 {r}}\\
\{\emptyset\}, & \text{otherwise. } \\
\end{cases}
\end{equation}

Combining \eqref{eq_codition_A22} and \eqref{eq_codition_A23} with intersections yields
\begin{equation}\label{eq_codition_A24}
\phi \in \begin{cases}
(\mathcal{F}_2, \pi], & \text{if } {r}^2{A} < B, \, \frac{-{r}^2{A} - B}{2 {r}} \leq l_{t,j}< -{r}{A} \\
(\mathcal{F}_2, \mathcal{F}_3], & \text{if }  {r}^2{A} < B,\,  l_{t,j} \geq -{r}{A}  \\
\{\emptyset\}, & \text{otherwise. } \\
\end{cases}
\end{equation}

Then, the intersections of \eqref{eq_codition_A24} and \eqref{eq_codition_A1_2} yield \eqref{eq_codition_A24}.

If $-{r}{A}+ l_{t,j} \cos \phi < 0$, then
\begin{equation} \label{eq_condition_A101}
\phi \in \begin{cases}
(\mathcal{F}_3, \pi],  &\text{if } l_{t,j} \geq -{r}{A}\\
\{\emptyset\}, & \text{otherwise. } \\
\end{cases}
\end{equation}

Then, combining \eqref{eq_condition_A101} and \eqref{eq_codition_A1_2} with intersections yields
\begin{equation}\label{eq_condition_A101_1_2}
\phi \in \begin{cases}
(\mathcal{F}_3, \pi] , & \text{if } {r}^2{A} \leq B, l_{t,j} \geq -{r}{A} \\
[\mathcal{F}_1, \pi], &\text{if } {r}^2{A} > B, l_{t,j}\geq \sqrt{B{A}} \\
\{\emptyset\}, & \text{otherwise. } \\
\end{cases}
\end{equation}

Therefore, the regions for $x_2 > {r}$ are given by the unions of \eqref{eq_codition_A24} and \eqref{eq_condition_A101_1_2}:

\begin{equation}\label{eq_condition_x_1_l_{r}}
\phi \in \begin{cases}
(\mathcal{F}_2, \pi], & \text{if } {r}^2{A} \leq B, \, \frac{-{r}^2{A} - B}{2 {r}} \leq l_{t,j}\\
[\mathcal{F}_1, \pi], &\text{if } {r}^2{A} > B, l_{t,j}\geq \sqrt{B{A}} \\
\{\emptyset\}, & \text{otherwise. } \\
\end{cases}
\end{equation}

Meanwhile, 
\begin{equation}\label {eq_sss}
\begin{split}
x_2\leq {r} \Rightarrow \sqrt{\left( l_{t,j} \cos \phi \right )^2 -  B{A}} \leq -{r}{A}+ l_{t,j} \cos \phi.\\
\end{split}
\end{equation}

If $-{r}{A}+ l_{t,j} \cos \phi\geq 0$, we have \eqref{eq_codition_A22}.

Assuming that \eqref{eq_codition_A1_con} and \eqref{eq_codition_A22} hold, solving \eqref{eq_sss} gives
\begin{equation}\label{eq_codition_A31}
\phi \in \begin{cases}
[0, \mathcal{F}_2], &  \text{if } l_{t,j} \geq \frac{-{r}^2{A} - B}{2 {r}} \\
[0, \pi] , &  \text{if } l_{t,j} < \frac{-{r}^2{A} - B}{2 {r}}. \\
\end{cases}
\end{equation}

Combining \eqref{eq_codition_A22} and \eqref{eq_codition_A31} with intersections yields 
\begin{equation}\label{eq_codition_A111}
\phi \in \begin{cases}
[0, \mathcal{F}_2], & \text{if } {r}^2{A} \leq B, l_{t,j} \geq  \frac{-{r}^2{A} - B}{2 {r}}  \\
[0, \mathcal{F}_3], & \text{if } {r}^2{A} > B,l_{t,j} \geq -{r}{A} \\
[0,\pi], & \text{if } {r}^2{A} \leq B, l_{t,j} < \frac{-{r}^2{A} - B}{2 {r}}\\
[0,\pi], & \text{if } {r}^2{A} > B, l_{t,j} < -{r}{A}\\
\{\emptyset\}, & \text{otherwise.}
\end{cases}
\end{equation}

The intersections of \eqref{eq_codition_A111} and \eqref{eq_codition_A1_2} are given by \eqref{eq_codition_A111_2}. As a result, when $0\leq x_1 \leq {r}$ and $x_2 \leq {r}$, we have \eqref{eq_codition_A111_2}. Meanwhile, when $0\leq x_1 \leq {r}$ and $x_2 > {r}$, we have
\eqref{eq_codition_A111_21}.

{\bf {For ${A}>0, B \geq 0$:}}
Assuming that \eqref{eq_codition_A1_con} holds, comparing $x_2$ with $0$ gives
\begin{equation}\label{eq_codition_2A1_bd_ex}
\begin{split}
\phi \in \left[\frac{\pi}{2}, \pi \right], \text{ for }  x_2 < 0; \text{ and } 
\phi \in \left[0,\frac{\pi}{2} \right], \text{ for } x_2 \geq 0.
\end{split}
\end{equation}

Assuming that \eqref{eq_codition_A1_con} holds, comparing $x_1$ with $0$ gives
\begin{equation}\label{eq_codition_2A1_bd}
\begin{split}
\phi \in \left[\frac{\pi}{2}, \pi \right], \text{ for }  x_1 < 0; \text{ and } 
\phi \in \left[0,\frac{\pi}{2} \right], \text{ for } x_1 \geq 0.
\end{split}
\end{equation}

Since $x_1$ must be larger than $0$ for a valid region to exist, $\phi \in \left[0,\frac{\pi}{2} \right]$ and \eqref{eq_codition_A1_1} stands. %

Comparing $x_2$ with ${r}$, we have
\begin{equation}\label{eq_condition_2A2_1}
\begin{split}
x_2\leq {r} \Rightarrow \sqrt{\left( l_{t,j} \cos \phi \right )^2 -  B{A}} \geq l_{t,j} \cos \phi - {r}{A}.\\
\end{split}
\end{equation}

If 
\begin{equation}\label{eq_ex_2}
l_{t,j} \cos \phi - {r}{A}\geq 0,
\end{equation}
then
\begin{equation}\label{eq_condition_2A2}
\phi \in \begin{cases}
[0, \mathcal{F}_3], & \text{if } l_{t,j} \geq {r}{A} \\
\{\emptyset\}, & \text{if }  l_{t,j} < {r}{A}. \\
\end{cases}
\end{equation}

Assuming that \eqref{eq_codition_A1_con}  and \eqref{eq_ex_2} hold, solving \eqref{eq_condition_2A2_1} gives
\begin{equation}\label{eq_codition_2A3}
\phi \in \begin{cases}
[0, \mathcal{F}_2], & \text{if } l_{t,j}\geq \frac{{r}^2{A} + B}{2 {r}} \\
\{\emptyset\}, & \text{if }  l_{t,j}<\frac{{r}^2{A} + B}{2 {r}}. \\
\end{cases}
\end{equation}

Then, combining \eqref{eq_condition_2A2}, \eqref{eq_codition_2A3} and \eqref{eq_codition_A1_1} with intersections yields

\begin{equation}\label{eq_codition_2A3_1}
\phi \in \begin{cases}
[0, \mathcal{F}_3], & \text{if } {r}^2{A} > B, l_{t,j} \geq {r}{A} \\
[0, \mathcal{F}_2], & \text{if } {r}^2{A} \leq B, l_{t,j}\geq \frac{{r}^2{A} + B}{2 {r}} \\
\{\emptyset\}, & \text{otherwise. } \\
\end{cases}
\end{equation}

If $l_{t,j} \cos \phi - {r}{A} < 0$, then

\begin{equation}\label{eq_codition_2A4}
\phi \in \begin{cases}
[0, \pi], & \text{if } l_{t,j} < {r}{A} \\
(\mathcal{F}_3, \pi], & \text{if }  l_{t,j} \geq {r}{A}. \\
\end{cases}
\end{equation}

Then, combining \eqref{eq_codition_2A4} and \eqref{eq_codition_A1_1} with intersections yields 

\begin{equation}\label{eq_codition_2A111_2}
\phi \in \begin{cases}
[0, \pi - \mathcal{F}_1], & \text{if } {r}^2{A}> B,  \sqrt{B{A}} \leq l_{t,j} < {r}{A}\\
(\mathcal{F}_3, \pi - \mathcal{F}_1], & \text{if } {r}^2{A} > B, {{r}{A}} \leq l_{t,j} \\
\{\emptyset\}, & \text{otherwise.}
\end{cases}
\end{equation}

The unions of \eqref{eq_codition_2A3_1} and \eqref{eq_codition_2A111_2} yield
\begin{equation}\label{eq_codition_2A111_3}
\phi \in \begin{cases}
[0, \mathcal{F}_2], & \text{if } {r}^2{A} \leq B, l_{t,j}\geq \frac{{r}^2{A} + B}{2 {r}} \\
[0, \pi - \mathcal{F}_1], & \text{if } {r}^2{A} > B, \sqrt{B{A}} \leq l_{t,j} \\
\{\emptyset\}, & \text{otherwise.}
\end{cases}
\end{equation}

Comparing ${r}$ with $x_1$, 
\begin{equation}\label{eq_codition_2A21}
\begin{split}
x_1\leq {r} \Rightarrow \sqrt{\left( l_{t,j} \cos \phi \right )^2 -  B{A}} \leq {r}{A}- l_{t,j} \cos \phi.\\
\end{split}
\end{equation}

If 
\begin{equation}\label{eq_ex_3}
{r}{A}- l_{t,j} \cos \phi\geq 0,\end{equation}
then
\begin{equation}\label{eq_codition_2A3_11}
\phi \in \begin{cases}
[0, \pi], & \text{if } l_{t,j} <{r}{A} \\
[\mathcal{F}_3, \pi], & \text{if }  l_{t,j}\geq {r}{A}. \\
\end{cases}
\end{equation}

Assuming that \eqref{eq_codition_A1_con} and \eqref{eq_ex_3} hold, solving \eqref{eq_codition_2A21} gives
\begin{equation}\label{eq_codition_2A23}
\phi \in \begin{cases}
[0,\pi],  &\text{if } l_{t,j}< \frac{{r}^2{A} + B}{2 {r}} \\
 \left[\mathcal{F}_2,\pi\right], & \text{if } l_{t,j}\geq \frac{{r}^2{A} + B}{2 {r}}. \\
\end{cases}
\end{equation}

Combining \eqref{eq_codition_2A23} and \eqref{eq_codition_2A3_11} with intersections yields
\begin{equation}\label{eq_codition_2A111}
\phi \in \begin{cases}
[0,\pi], & \text{if } {r}^2{A} < B, l_{t,j} < {{r}{A}}\\
[0,\pi], & \text{if } {r}^2{A} \geq B, l_{t,j} < \frac{{r}^2{A} + B}{2 {r}}\\
[\mathcal{F}_3, \pi], & \text{if } {r}^2{A} < B, {{r}{A}  \leq l_{t,j}}\\
[\mathcal{F}_2, \pi] , & \text{if } {r}^2{A} \geq B, \frac{{r}^2{A} + B}{2 {r}} \leq l_{t,j} \\
\{\emptyset\}, & \text{otherwise.}
\end{cases}
\end{equation}

Then, combining \eqref{eq_codition_2A111} and \eqref{eq_codition_A1_1} with intersections yields
\begin{equation}\label{eq_codition_2A111_9}
\phi \in \begin{cases}
[\mathcal{F}_2, \pi - \mathcal{F}_1], & \text{if } {r}^2{A} \geq B, l_{t,j} \geq \frac{{r}^2{A} + B}{2 {r}}\\
[0, \pi - \mathcal{F}_1], & \text{if } {r}^2{A} \geq B, \sqrt{B{A}} \leq l_{t,j} < \frac{{r}^2{A} + B}{2 {r}}\\
\{\emptyset\}, & \text{otherwise.}
\end{cases}
\end{equation}
\begin{equation}\label{eq_codition_2A21_1}
\begin{split}
x_1> {r} \Rightarrow \sqrt{\left( l_{t,j} \cos \phi \right )^2 -  B{A}} > {r}{A}- l_{t,j} \cos \phi.\\
\end{split}
\end{equation}

If 
\begin{equation}\label{eq_ex_1}
{r}{A}- l_{t,j} \cos \phi\geq 0,
\end{equation} we have \eqref{eq_codition_2A3_11}.

Assuming that \eqref{eq_codition_A1_con} and \eqref{eq_ex_1} hold, solving \eqref{eq_codition_2A21_1} gives  
\begin{equation}\label{eq_codition_2A23_1}
\phi \in \begin{cases}
[0, \mathcal{F}_2), & \text{if } l_{t,j}\geq \frac{{r}^2{A} + B}{2 {r}} \\
\{\emptyset\}, & \text{otherwise.}
\end{cases}
\end{equation}

Combining \eqref{eq_codition_2A3_11} and \eqref{eq_codition_2A23_1} with intersections yields
\begin{equation}\label{eq_codition_2A23_2}
\phi \in \begin{cases}
[\mathcal{F}_3, \mathcal{F}_2), & \text{if } {r}^2{A} > B, l_{t,j}\geq {r}{A} \\
[0, \mathcal{F}_2), & \text{if } {r}^2{A} > B, \frac{{r}^2{A} + B}{2 {r} } \leq l_{t,j}< {r}{A} \\
\{\emptyset\}, & \text{otherwise.}
\end{cases}
\end{equation}

If ${r}{A}- l_{t,j} \cos \phi< 0$, we have
\begin{equation}\label{eq_codition_2A23_11}
\phi \in \begin{cases}
[0, \mathcal{F}_3) ,  &\text{if } l_{t,j}\geq {{r}{A}} \\
\{\emptyset\}, & \text{otherwise.}
\end{cases}
\end{equation}

The unions of \eqref{eq_codition_2A23_2} and \eqref{eq_codition_2A23_11} yield
\begin{equation}\label{eq_codition_2A23_21}
\phi \in \begin{cases}
[0, \mathcal{F}_2), & \text{if } {r}^2{A} > B, l_{t,j}\geq  \frac{{r}^2{A} + B}{2 {r} } \\
[0, \mathcal{F}_3),  &\text{if } {r}^2{A} \leq  B, l_{t,j}\geq {{r}{A}} \\
\{\emptyset\}, & \text{otherwise.}
\end{cases}
\end{equation}

Then, combining \eqref{eq_codition_2A23_21} and \eqref{eq_codition_A1_1} with intersections yields
\begin{equation}\label{eq_codition_2A23_211}
\phi \in \begin{cases}
[0, \mathcal{F}_2), & \text{if } {r}^2{A} > B, l_{t,j}\geq \frac{{r}^2{A} + B}{2 {r} } \\
[0, \pi - \mathcal{F}_1] ,  &\text{if } {r}^2{A} \leq  B, l_{t,j}\geq {\sqrt{B{A}}} \\
\{\emptyset\}, & \text{otherwise.}
\end{cases}
\end{equation}

The feasible regions of $\phi$, where $x_2 \leq {r}$ and $x_1 \leq {r}$, are the intersections of \eqref{eq_codition_2A111_3} and \eqref{eq_codition_2A23_211}. As a result,
\begin{equation}
\phi \in \begin{cases}
[\mathcal{F}_2, \pi - \mathcal{F}_1], & \text{if } {r}^2{A} \geq B, l_{t,j} \geq \frac{{r}^2{A} + B}{2 {r}}\\
[0, \pi - \mathcal{F}_1], & \text{if } {r}^2{A} > B, \sqrt{B{A}} \leq l_{t,j} < \frac{{r}^2{A} + B}{2 {r}}\\
\{\emptyset\}, & \text{otherwise.}
\end{cases}
\end{equation}

The feasible regions of $\phi$, where $x_2 \leq {r}$ and $x_1 > {r}$, are the intersections of \eqref{eq_codition_2A111_3} and \eqref{eq_codition_2A111_9}. As a result,
\begin{equation} 
\phi \in \begin{cases}
[0,  \mathcal{F}_2],  &\text{if } {r}^2{A} \leq  B, \frac{{r}^2{A} + B}{2 {r}} \leq l_{t,j}  \\
[0,  \mathcal{F}_2),  &\text{if } {r}^2{A} > B, \frac{{r}^2{A} + B}{2 {r}} \leq l_{t,j}  \\
\{\emptyset\}, & \text{otherwise.}
\end{cases}
\end{equation}
As a result, when $x_2 \leq {r}$ and $x_1 \leq {r}$, we have \eqref{eq_lemma_1_3}. Meanwhile, when $x_2 \leq {r}$ and $x_1 > {r}$, we have
\eqref{eq_lemma_1_4}.

{\bf{ For ${A}<0, B > 0$:}}
Comparing $x_1$ with $0$ gives 
\begin{equation}
\begin{split}
\phi \in \left[0, \pi \right],  \text{ for } x_1 < 0; \text{ and }
\phi = \emptyset,  \text{ for }x_1 \geq 0.
\end{split}
\end{equation}

Comparing $x_2$ with $0$ gives 
\begin{equation}
\begin{split}
\phi = \emptyset, \text{ for } x_2 < 0;  \text{ and }
\phi \in \left[0, \pi \right], \text{ for } x_2 \geq 0.
\end{split}
\end{equation}

Comparing $x_2$ with ${r}$, 
\begin{equation}\label{eq_condition_3A21}
\begin{split}
x_2 > {r} \Rightarrow \sqrt{\left( l_{t,j} \cos \phi \right )^2 -  B{A}} > l_{t,j} \cos \phi - {r}{A}.\\
\end{split}
\end{equation}

If $l_{t,j} \cos \phi - {r}{A} < 0$, then
\begin{equation}\label{eq_condition_2A21}
\phi \in \begin{cases}
(\mathcal{F}_3, \pi] , & \text{if } l_{t,j} \geq -{r}{A} \\
\{\emptyset\}, & \text{if }  l_{t,j} < -{r}{A}. \\
\end{cases}
\end{equation}

If 
\begin{equation}\label{eq_ex_6}
l_{t,j} \cos \phi - {r}{A} \geq 0,  
\end{equation}
then
\begin{equation}\label{eq_condition_3A2}
\phi \in \begin{cases}
[0, \mathcal{F}_3], & \text{if } l_{t,j} \geq -{r}{A} \\
[0, \pi], & \text{if }  l_{t,j} < -{r}{A}. \\
\end{cases}
\end{equation}

Assuming that \eqref{eq_codition_A1_con} and \eqref{eq_ex_6} hold, solving \eqref{eq_condition_3A21} gives
\begin{equation}\label{eq_codition_3A23}
\phi \in \begin{cases}
[0,\pi],  &\text{if } -{r}^2{A} <B, l_{t,j}< \frac{{r}^2{A} + B}{2 {r}} \\
 (\mathcal{F}_2, \pi], & \text{if } -{r}^2{A} \geq B, l_{t,j}\geq \frac{-{r}^2{A} - B}{2 {r}} \\
(\mathcal{F}_2, \pi], & \text{if } -{r}^2{A} < B, l_{t,j}\geq \frac{{r}^2{A} + B}{2 {r}} \\
\{\emptyset\}, & \text{otherwise.}
\end{cases}
\end{equation}

Then, combining \eqref{eq_condition_3A2} and \eqref{eq_codition_3A23} with intersections yields
\begin{equation} \SMALL \label{eq_codition_3A23_1}
\phi \in \begin{cases}
[0,\pi],  &\text{if } -{A} < \frac{ B}{{r}^2}, -{A} \geq \frac{B}{3{r}^2}, l_{t,j}< \frac{{r}^2{A} + B}{2 {r}} \\
[0,\pi],  &\text{if } -{A} < \frac{ B}{{r}^2}, -{A} <  \frac{B}{3{r}^2}, l_{t,j}< -{r}{A} \\
(\mathcal{F}_2, \pi], & \text{if } -{A} \geq \frac{B}{{r}^2},  \frac{-{r}^2{A} - B}{2 {r}} \leq l_{t,j} < -{r}{A} \\
(\mathcal{F}_2, \pi], & \text{if } -{A} < \frac{ B}{{r}^2}, -{A} > \frac{B}{3{r}^2}, \frac{{r}^2{A} + B}{2 {r}} \leq l_{t,j} < -{r}{A} \\
[0, \mathcal{F}_3], & \text{if } -{A} < \frac{ B}{{r}^2}, -{A} <  \frac{B}{3{r}^2},   -{r}{A} \leq l_{t,j} < \frac{{r}^2{A} + B}{2 {r}}\\
 (\mathcal{F}_2, \mathcal{F}_3], & \text{if } -{A} \geq \frac{B}{{r}^2}, l_{t,j}\geq -{r}{A} \\
(\mathcal{F}_2, \mathcal{F}_3], & \text{if } -{A} < \frac{ B}{{r}^2}, -{A} \geq \frac{B}{3{r}^2}, l_{t,j}\geq -{r}{A} \\
(\mathcal{F}_2, \mathcal{F}_3], & \text{if } -{A} < \frac{ B}{{r}^2}, -{A} <  \frac{B}{3{r}^2}, l_{t,j}\geq \frac{{r}^2{A} + B}{2 {r}} \\
\{\emptyset\}, & \text{otherwise.}
\end{cases}
\end{equation}

The unions of \eqref{eq_condition_2A21} and \eqref{eq_codition_3A23_1} yield
\begin{equation}\label{Eq_union_001}
\phi \in \begin{cases}
[0,\pi],  &\text{if } -{r}^2{A} < B, l_{t,j}< \frac{{r}^2{A} + B}{2 {r}} \\
(\mathcal{F}_2, \pi], & \text{if }  l_{t,j}\geq  \frac{\left |{r}^2{A} + B\right|}{2 {r}}  \\
\{\emptyset\}, & \text{otherwise.}
\end{cases}
\end{equation}

Meanwhile,
\begin{equation}\label{eq_condition_3A211}
\begin{split}
x_2 \leq {r} \Rightarrow \sqrt{\left( l_{t,j} \cos \phi \right )^2 -  B{A}} \leq l_{t,j} \cos \phi - {r}{A}.\\
\end{split}
\end{equation}

If $l_{t,j} \cos \phi - {r}{A} \geq 0$, 
then
\begin{equation}\label{eq_condition_3A2111}
\phi \in \begin{cases}
[0, \mathcal{F}_3], & \text{if } l_{t,j} \geq -{r}{A} \\
[0, \pi], & \text{if }  l_{t,j} < -{r}{A}. \\
\end{cases}
\end{equation}

Assuming that \eqref{eq_codition_A1_con} hold, solving \eqref{eq_condition_3A211} gives
\begin{equation}\label{eq_codition_3A2311}
\phi \in \begin{cases}
[0,\pi],  &\text{if } -{r}^2{A} > B, l_{t,j}< \frac{-{r}^2{A} - B}{2 {r}} \\
[0, \mathcal{F}_2], & \text{if } l_{t,j}\geq \frac{\left| -{r}^2{A} - B\right|}{2 {r}} \\
\{\emptyset\}, & \text{otherwise.}
\end{cases}
\end{equation}
The intersections of \eqref{eq_condition_3A2111} and \eqref{eq_codition_3A2311} are given by \eqref{eq_codition_3A2311}.
Hence, the feasible regions for $x_2>{r}$ or $0\leq x_2\leq {r}$ are given by \eqref{Eq_union_001} and \eqref{eq_codition_3A2311}, respectively.
As a result, when $x_2>{r}$, we have \eqref{eq_lemma_1_5}. Meanwhile, when $0\leq x_2\leq {r}$, we have
\eqref{eq_lemma_1_6}.

{\bf For ${A}>0, B < 0$:}
Comparing $x_1$ with $0$ gives
\begin{equation}
\begin{split}
\phi \in \left[0, \pi \right], \text{ for } x_1 \geq 0; \text{ and } \phi=\emptyset,  \text{ for } x_1 < 0.
\end{split}
\end{equation}

Comparing $x_2$ with $0$ gives
\begin{equation}
\begin{split}
\phi =\emptyset, \text{ for } x_2 \geq 0; \text{ and }
\phi \in \left[0, \pi \right]\text{ for } \,x_2 < 0.
\end{split}
\end{equation}

Comparing $x_1$ with ${r}$, 
\begin{equation}\label{eq_condition_3A211_g}
\begin{split}
x_1 > {r} \Rightarrow \sqrt{\left( l_{t,j} \cos \phi \right )^2 -  B{A}} >  {r}{A} - l_{t,j} \cos \phi .\\
\end{split}
\end{equation}

If ${r}{A} - l_{t,j} \cos \phi < 0$, then
\begin{equation}\label{eq_condition_3A2_1_1}
\phi \in \begin{cases}
[0, \mathcal{F}_3), & \text{if } l_{t,j} \geq {r}{A} \\
\{\emptyset\}, & \text{if } l_{t,j} < {r}{A}. \\
\end{cases}
\end{equation}

If 
\begin{equation}\label{eq_ex_8}
{r}{A} - l_{t,j} \cos \phi \geq 0, 
\end{equation}
then
\begin{equation}\label{eq_condition_3A211_1}
\phi \in \begin{cases}
[\mathcal{F}_3, \pi], & \text{if } l_{t,j} \geq {r}{A} \\
[0, \pi], & \text{if }  l_{t,j} < {r}{A}. \\
\end{cases}
\end{equation}

Assuming that \eqref{eq_codition_A1_con} and \eqref{eq_ex_8} hold, solving \eqref{eq_condition_3A211_g} gives
\begin{equation}\label{eq_codition_3A23_9}
\phi \in \begin{cases}
[0,\pi],  &\text{if } -{r}^2{A} > B, l_{t,j}< \frac{-{r}^2{A} - B}{2 {r}} \\
[0, \mathcal{F}_2], & \text{if } -{r}^2{A} \leq B, l_{t,j} \geq \frac{{r}^2{A} + B}{2 {r}} \\
[0, \mathcal{F}_2], & \text{if } -{r}^2{A} > B, l_{t,j}\geq \frac{-{r}^2{A} - B}{2 {r}} \\
\{\emptyset\}, & \text{otherwise.}
\end{cases}
\end{equation}

Then, combining \eqref{eq_condition_3A211_1} and \eqref{eq_codition_3A23_9} with intersections yields
\begin{equation}\SMALL \label{eq_codition_3A23_11_1}
\phi \in \begin{cases}
[0,\pi],  &\text{if } -{A} > \frac{ B}{{r}^2}, -{A} \geq \frac{B}{3{r}^2}, l_{t,j}< {r}{A} \\
[0,\pi],  &\text{if } -{A} > \frac{ B}{{r}^2}, -{A} <  \frac{B}{3{r}^2}, l_{t,j}< \frac{-{r}^2{A} - B}{2 {r}} \\
[0, \mathcal{F}_2], & \text{if } -{A} \leq \frac{B}{{r}^2}, \frac{{r}^2{A} + B}{2 {r}}\leq l_{t,j} < {r}{A} \\
[0, \mathcal{F}_2], & \text{if } -{A} > \frac{ B}{{r}^2}, -{A} <  \frac{B}{3{r}^2}, \frac{-{r}^2{A} - B}{2 {r}} \leq l_{t,j} < {r}{A}\\
[\mathcal{F}_3, \pi], & \text{if } -{A} > \frac{ B}{{r}^2}, -{A} > \frac{B}{3{r}^2}, {r}{A} \leq l_{t,j} < \frac{-{r}^2{A} - B}{2 {r}}\\
[\mathcal{F}_3, \mathcal{F}_2], & \text{if } -{A} \leq \frac{B}{{r}^2}, l_{t,j} \geq {r}{A} \\
[\mathcal{F}_3, \mathcal{F}_2], & \text{if } -{A} > \frac{ B}{{r}^2}, -{A} \geq \frac{B}{3{r}^2}, l_{t,j}\geq \frac{-{r}^2{A} - B}{2 {r}} \\
[\mathcal{F}_3, \mathcal{F}_2], & \text{if } -{A} > \frac{ B}{{r}^2}, -{A} <  \frac{B}{3{r}^2}, l_{t,j}\geq {r}{A} \\
\{\emptyset\}, & \text{otherwise.}
\end{cases}
\end{equation}
The unions of \eqref{eq_condition_3A2_1_1} and \eqref{eq_codition_3A23_11_1} yield
\begin{equation}\label{Eq_union_002}
\phi \in \begin{cases}
[0,\pi],  &\text{if } -{r}^2{A} > B, l_{t,j}< \frac{-{r}^2{A} - B}{2 {r}} \\
[0, \mathcal{F}_2], & \text{if } l_{t,j}\geq \frac{|{r}^2{A} + B |}{2 {r}} \\
\{\emptyset\}, & \text{otherwise.}
\end{cases}
\end{equation}

Meanwhile,
\begin{equation}\label{eq_condition_3A21112}
\begin{split}
x_1 \leq {r} \Rightarrow \sqrt{\left( l_{t,j} \cos \phi \right )^2 -  B{A}} \leq {r}{A} - l_{t,j} \cos \phi .\\
\end{split}
\end{equation}

If 
\begin{equation} \label{eq_ex_7}
{r}{A} - l_{t,j} \cos \phi \geq 0, 
\end{equation}
then
\begin{equation}\label{eq_condition_3A21111}
\phi \in \begin{cases}
[\mathcal{F}_3, \pi], & \text{if } l_{t,j} \geq {r}{A} \\
[0, \pi], & \text{if }  l_{t,j} < {r}{A}. \\
\end{cases}
\end{equation}

Assuming that \eqref{eq_codition_A1_con} and \eqref{eq_ex_7} hold, solving \eqref{eq_condition_3A21112} gives
\begin{equation}\label{eq_codition_3A23111}
\phi \in \begin{cases}
[0,\pi],  &\text{if } -{r}^2{A} < B, l_{t,j}< \frac{{r}^2{A} + B}{2 {r}} \\
[\mathcal{F}_2, \pi], & \text{if } l_{t,j}\geq \frac{|{r}^2{A} + B|}{2 {r}} \\
\{\emptyset\}, & \text{otherwise.}
\end{cases}
\end{equation}
The intersections of \eqref{eq_condition_3A21111} and \eqref{eq_codition_3A23111} is \eqref{eq_codition_3A23111}.
Hence, the feasible regions for $x_1>{r}$ or $0\leq x_1\leq {r}$ are given by \eqref{Eq_union_002} and \eqref{eq_codition_3A23111}, respectively.
As a result of the above discussions, when $x_1>{r}$, we have \eqref{eq_lemma_1_7}. Meanwhile, when $0\leq x_1\leq {r}$, we have
\eqref{eq_lemma_1_8}.

{\bf For ${A}= 0$:} The corresponding solution to $1+\frac{(H^2+l_{t,e}^2)^\frac{-\alpha}{2}\eta  P_s}{\left(l_{t,j}^2+l_{t,e}^2-2 l_{t,j} l_{t,e}\cos(\phi )\right)^\frac{-\alpha}{2}P_j} \leq y$  is given by
$
{\bf{{ \cos \phi}} }\geq  \frac{ l_{t,j}^2 -H^2}{2 l_{t,j}  l_{t,e}}.
$ Thus, we have
\begin{equation}\label{eq_new_G1}
\phi \in \begin{cases}
\left[0, {\cos^{-1}} \left(\frac{ l_{t,j}^2 -H^2}{2 l_{t,j}  l_{t,e}}\right) \right], & \text{if }   \frac{\left | l_{t,j}^2 -H^2\right|}{2 l_{t,j}  }   \leq l_{t,e} \\
[0,\pi],  &\text{if } \frac{ -l_{t,j}^2 +H^2}{2 l_{t,j}  } >l_{t,e}  \\
\{\emptyset\}, & \text{otherwise.}
\end{cases}
\end{equation}

Comparing $l_{t,e}$ from \eqref{eq_new_G1} with ${r}$, we have \eqref{eq_new_region_A1} - \eqref{eq_new_region_A3}.

\end{IEEEproof}


\bibliographystyle{IEEEtran}
\bibliography{IEEEabrv,papers}

\end{document}